\documentclass[11pt]{article} 
\usepackage[left=2.5cm, right=2.5cm, top=1.785cm, bottom=2.0cm]{geometry}

\usepackage[utf8]{inputenc}
\usepackage{amsmath}
\usepackage{amsfonts}
\usepackage{amssymb}
\usepackage{mathtools}
\usepackage{listings}
\usepackage{mathrsfs}
\usepackage{times}
\usepackage{float}
\usepackage{color}
\usepackage{setspace}
\usepackage{color,soul}

\usepackage{amsmath,amsfonts,amsthm,eucal}
\usepackage{enumerate}
\usepackage{color}
\usepackage{tabularx}
\usepackage{siunitx}
\usepackage{empheq}
\usepackage[tableposition=below]{caption}
\usepackage[
pdfstartview=XYZ,
bookmarks=true,
colorlinks=true,
linkcolor=blue,
urlcolor=blue,
citecolor=blue,
pdftex,
bookmarks=true,
linktocpage=true,   
hyperindex=true
]{hyperref}

\usepackage{lipsum}

\usepackage[pdftex]{graphicx}
\usepackage{epstopdf}
\usepackage{booktabs} 
\usepackage{tikz,mathpazo}
\usetikzlibrary{shapes.geometric, arrows}
\usepackage{caption}

\usepackage{longtable}
\usepackage{adjustbox}
\usepackage{framed}

\usepackage[colorinlistoftodos]{todonotes}

\usepackage{listings} 
\usepackage{showexpl} 
\usepackage{color} 
\usepackage{amsthm} 
\usepackage{alltt}

\usepackage[ruled,vlined]{algorithm2e}
\usepackage{algpseudocode}
\usepackage{amssymb}
\usepackage{amsmath}
\usepackage{amsfonts}
\usepackage{booktabs}
\usepackage{enumerate}
\usepackage{fancyhdr}
\usepackage{float}
\usepackage{graphicx}
\usepackage{lastpage}
\usepackage{mathrsfs}
\usepackage{hyperref}
\hypersetup{
    colorlinks=true,
    linkcolor=blue,
    filecolor=magenta,      
    urlcolor=cyan,
}

\usepackage{appendix}
\usepackage{tikz}
\usetikzlibrary{decorations.text}
\usetikzlibrary{shapes,arrows}
\usepackage{pgfplots}
\usepackage{pgfplotstable}
\pgfplotsset{compat=newest}

\usetikzlibrary{calc}
\usetikzlibrary{spy}
\usetikzlibrary{positioning}
\usepackage{pdftexcmds}
\usetikzlibrary{external}
\usetikzlibrary{positioning}
\usetikzlibrary{arrows}
\usetikzlibrary{decorations.markings}
\usetikzlibrary{decorations.text}
\usetikzlibrary{backgrounds}
\usetikzlibrary{spy}
\usetikzlibrary{calc,patterns,decorations.pathmorphing,decorations.markings}
\usetikzlibrary{decorations.pathreplacing}
\usepgfplotslibrary{patchplots}

\usepackage{chngcntr}
\usepackage{etoolbox}
\usepackage{lipsum}

\usepackage{bm}

\makeatletter
\providecommand*{\input@path}{}
\g@addto@macro\input@path{{./}}
\makeatother

\makeatletter

\def\pgfplotstableread@openfile{%
    \def\pgfplotstable@loc@TMPa{\pgfutil@in@{ }}%
    \expandafter\pgfplotstable@loc@TMPa\expandafter{\pgfplotstableread@filename}%
    \ifpgfutil@in@
        \t@pgfplots@toka=\expandafter{\pgfplotstableread@filename}%
        \edef\pgfplotstableread@filename{\pgfplots@dquote\the\t@pgfplots@toka\pgfplots@dquote}%
    \fi
    \let\pgfplotstableread@old@crcr=\\%
    \def\\{\string\\}
    \openin\r@pgfplots@reada=\csname pgfk@/pgfplots/table file path\endcsname\pgfplotstableread@filename.tex
    \ifeof\r@pgfplots@reada
        \openin\r@pgfplots@reada=\csname pgfk@/pgfplots/table file path\endcsname\pgfplotstableread@filename\relax
    \else
        \pgfplots@warning{%
            You requested to open table '\pgfplotstableread@filename', but there is also a '\pgfplotstableread@filename.tex'. 
            TeX will automatically append the suffix '.tex', so I will now open '\pgfplotstableread@filename.tex'.
            Please make sure you don't accidentally load TeX files - this may produce unrecoverable errors.}%
        \closein\r@pgfplots@reada
        \openin\r@pgfplots@reada=\pgfplotstableread@filename\relax
    \fi
    \ifeof\r@pgfplots@reada
        \pgfplotsthrow{no such table file}{\pgfplots@loc@TMPa}{\pgfplotstableread@filename}{Could not read table file '\csname pgfk@/pgfplots/table file path\endcsname\pgfplotstableread@filename'. In case you intended to provide inline data: maybe TeX screwed up your end-of-lines? Try `row sep=crcr' and terminate your lines with `\string\\' (refer to the pgfplotstable manual for details)}\pgfeov%
        \global\let\pgfplotstable@colnames@glob=\pgfplots@loc@TMPa
        \def\pgfplotstableread@ready{0}%
    \fi
    \pgfplots@logfileopen{\pgfplotstableread@filename}%
    \let\\=\pgfplotstableread@old@crcr
}

\makeatother

\pgfplotscreateplotcyclelist{blue to red}{%
  color=blue\\%
  color=red!10!blue\\%
  color=red!20!blue\\%
  color=red!30!blue\\%
  color=red!40!blue\\%
  color=red!50!blue\\%
  color=red!60!blue\\%
  color=red!70!blue\\%
  color=red!80!blue\\%
  color=red!90!blue\\%
  color=red\\%
}
\pgfplotscreateplotcyclelist{red to blue}{%
  color=red\\%
  color=red!90!blue\\%
  color=red!80!blue\\%
  color=red!70!blue\\%
  color=red!60!blue\\%
  color=red!50!blue\\%
  color=red!40!blue\\%
  color=red!30!blue\\%
  color=red!20!blue\\%
  color=red!10!blue\\%
  color=blue\\%
}

\pgfplotscreateplotcyclelist{red to blue fast}{%
  color=red\\%
  color=red!80!blue\\%
  color=red!60!blue\\%
  color=red!40!blue\\%
  color=red!20!blue\\%
  color=blue\\%
}

\pgfplotscreateplotcyclelist{2red to blue}{%
  color=red\\%
  color=red\\%
  color=red!90!blue\\%
  color=red!90!blue\\%
  color=red!80!blue\\%
  color=red!80!blue\\%
  color=red!70!blue\\%
  color=red!70!blue\\%
  color=red!60!blue\\%
  color=red!60!blue\\%
  color=red!50!blue\\%
  color=red!50!blue\\%
  color=red!40!blue\\%
  color=red!40!blue\\%
  color=red!30!blue\\%
  color=red!30!blue\\%
  color=red!20!blue\\%
  color=red!20!blue\\%
  color=red!20!blue\\%
  color=red!10!blue\\%
  color=blue\\%
  color=blue\\%
}

\pgfplotsset{discard if/.style 2 args={x filter/.code={\ifnum\thisrow{#1}=#2\else\fi}}}

\usepackage{bm}

\definecolor{deepblue}{rgb}{0,0,1}
\definecolor{deepred}{rgb}{1,0,0}
\definecolor{deepgreen}{rgb}{0,0.8,0}

\newcommand\pythonstyle{\lstset{
backgroundcolor=\color{white},
language=Python,
xleftmargin=0.05\textwidth,
xrightmargin=0.05\textwidth,
basicstyle=\footnotesize\ttfamily,
otherkeywords={def,self,add,reshape,zeros_like,isclose,vjp},             
keywordstyle=\color{deepblue},
emph={MyClass,__init__,import,as,return},          
emphstyle=\color{deepred},    
stringstyle=\color{deepgreen},
frame=single,                         
showstringspaces=false            %
}}

\lstnewenvironment{python}[1][]
{
\pythonstyle
\lstset{#1}
}
{}

\newcommand{\vepsb}{\boldsymbol \varepsilon}

\newcommand{\lambdab}{{\boldsymbol \lambda}}

\newcommand{\phib}{\boldsymbol \phi}

\newcommand{\cs}{\boldsymbol \sigma}

\newcommand{\thetab}{\boldsymbol \theta}

\newcommand{\Ab}{{\boldsymbol A}}

\newcommand{\Cb}{{\boldsymbol C}}

\newcommand{\Fb}{{\boldsymbol F}}

\newcommand{\Ib}{{\boldsymbol I}}
\newcommand{\Kb}{{\boldsymbol K}}

\newcommand{\Pb}{{\boldsymbol P}}

\newcommand{\Rb}{{\boldsymbol R}}

\newcommand{\Ub}{{\boldsymbol U}}

\newcommand{\bb}{{\boldsymbol b}}

\newcommand{\fb}{{\boldsymbol f}}

\newcommand{\nb}{{\boldsymbol n}}

\newcommand{\rb}{{\boldsymbol r}}
\newcommand{\sbf}{{\boldsymbol s}}
\newcommand{\tb}{{\boldsymbol t}}
\newcommand{\ub}{{\boldsymbol u}}
\newcommand{\vb}{{\boldsymbol v}}

\newcommand{\x}{{\boldsymbol x}}

\newcommand{\X}{{\boldsymbol X}}

\def\gJ{{\mathcal{J}}}

\def\sR{{\mathbb{R}}}

\usepackage{amsmath}

\usepackage{nicefrac}
\usepackage{multirow}
\usepackage{hhline}
\usepackage{lineno}
\usepackage{import}
\usepackage{epstopdf}
\usepackage{subcaption}
\usepackage{mathtools}
\usepackage{transparent}

\usepackage{cancel}
\usepackage{empheq}
\usepackage{booktabs}
\usepackage{cleveref}
\usepackage{soul}
\usepackage{courier}

\SetKwInput{KwInput}{Input}
\SetKwInput{KwOutput}{Output}

\definecolor{tbf}{RGB}{255,0,0} 
\definecolor{txue}{RGB}{0,0,255}

\usepackage{authblk}

\title{\Large JAX-FEM: A differentiable GPU-accelerated 3D finite element solver for automatic inverse design and mechanistic data science} 

\begin{document}

\author[1]{\normalsize Tianju Xue}
\author[1]{\normalsize Shuheng Liao}
\author[2]{\normalsize Zhengtao Gan}
\author[1]{\normalsize Chanwook Park}
\author[1]{\normalsize Xiaoyu Xie}
\author[1]{\normalsize Wing Kam Liu}
\author[1]{\normalsize Jian Cao
\footnote{\textit{jcao@northwestern.edu}  (corresponding author)}}
\affil[1]{\footnotesize Department of Mechanical Engineering, Northwestern University, Evanston, IL 60208,USA}
\affil[2]{\footnotesize Department of Aerospace and Mechanical Engineering, The University of Texas at El Paso, El Paso, TX 79968, USA}

\date{}
\maketitle

\vspace{-30pt}

\begin{abstract}
This paper introduces JAX-FEM, an open-source differentiable finite element method (FEM) library.
Constructed on top of Google JAX, a rising machine learning library focusing on high-performance numerical computing, JAX-FEM is implemented with pure Python while scalable to efficiently solve problems with moderate to large sizes.
For example, in a 3D tensile loading problem with 7.7 million degrees of freedom, JAX-FEM with GPU achieves  around 10$\times$ acceleration compared to a commercial FEM code depending on platform. 
Beyond efficiently solving forward problems, JAX-FEM employs the automatic differentiation technique so that inverse problems are solved in a fully automatic manner without the need to manually derive sensitivities.
Examples of 3D topology optimization of nonlinear materials are shown to achieve optimal compliance.
Finally, JAX-FEM is an integrated platform for machine learning-aided computational mechanics.
We show an example of data-driven multi-scale computations of a composite material where JAX-FEM provides an all-in-one solution from microscopic data generation and model training to macroscopic FE computations.
The source code of the library and these examples are shared with the community to facilitate computational mechanics research. 

\end{abstract}

\section{Introduction}

Research in computational science and engineering involving partial differential equations (PDEs) has long focused on developing efficient numerical algorithms.
Yet the overall efficiency of PDE-based computational analysis depends not only on the smart use of computer hardware, but also on the efficient use of human resources~\cite{kamensky2019tigar}.

The finite element method (FEM)~\cite{hughes2012finite} is one of the most powerful approaches for numerical solutions to PDEs that appear in structural analysis, heat transfer, fluid flow, electromagnetic potential, etc.
This paper proposes a library called \texttt{JAX-FEM} that aims at automating the finite element analysis workflows and enhancing human productivity.
\texttt{JAX-FEM} is built on \texttt{JAX}~\cite{jax2018github}, a library for high-performance numerical computing and machine learning research.
Besides its success in machine learning applications, \texttt{JAX} has proven to be a powerful building block for high-performance scientific simulations, including computational fluid dynamics~\cite{kochkov2021machine,bezgin2023jax}, structural dynamics of metamaterials~\cite{xue2022learning}, molecular dynamics~\cite{schoenholz2020jax}, phase-field simulation of microstructure evolution~\cite{xue2022physics}, etc.
We emphasize the following three features that differentiate \texttt{JAX-FEM} from other FEM libraries:
\begin{enumerate}
    \item Efficient solution to forward PDE with GPU acceleration;
    \item Differentiable simulation for automatic inverse design;
    \item Seamless integration with machine learning.
\end{enumerate}
In the following paragraphs, we introduce the background and motivation for the three features in detail.

\paragraph{Feature 1} 
Most existing FEM libraries (e.g., the \texttt{Dealii}~\cite{bangerth2007deal} library) are implemented with compiled languages like \texttt{C/C++} or \texttt{Fortran}. 
\texttt{JAX-FEM} is implemented with pure \texttt{Python}, a higher-level language known for its dynamic nature and flexibility. 
Due to the \texttt{XLA} (Accelerated Linear Algebra) backend of \texttt{JAX}, \texttt{JAX-FEM} has a highly competitive performance, especially when GPU is available. 
Therefore, our \texttt{Python} frontend provides both users and developers with fast production experience without condemning them to only small-sized problems.
Another benefit of \texttt{JAX-FEM} is automatic linearization for nonlinear problems.
Many FEM libraries require users to derive the linearized incremental form in Newton's method as input to the program, while \texttt{JAX-FEM} works directly with the weak form and performs the linearization using 
automatic differentiation (AD)~\cite{griewank2008evaluating} for the user.
While the AD technique is well known to be the workhorse of deep learning~\cite{lecun2015deep}, its potential in scientific applications has just gained more attention in recent years~\cite{vigliotti2021automatic,lindsay2021automatic,mozaffar4160375differentiable}.

\paragraph{Feature 2} Inverse design problems are of great interest in various engineering applications. 
Solving inverse problems are more challenging and computationally demanding due to the need to iteratively solve the forward problems.
Mathematically, inverse problems can often be formulated as PDE-constrained optimization problems~\cite{rees2010optimal}.
Successfully computing the ``sensitivity'' (gradient of the objective function to design parameters) is key to gradient-based optimization algorithms.
The derivation of the sensitivity, however, can be quite non-trivial particularly when the forward problem involves complicated nonlinear constitutive relations~\cite{van2005review}.
Based on automatic differentiation, \texttt{JAX-FEM} computes the sensitivity in a fully automatic manner, freeing the users from deriving the sensitivities by hand. 
The dimension of design parameters is usually much larger than the design objective, hence the the adjoint method~\cite{errico1997adjoint,cao2003adjoint} is used for efficiency.

\paragraph{Feature 3} The ever-increasing interest in data-driven computational mechanics in recent years has posed a strong need for an integrated platform with unified solutions.
For example, machine learning-based constitutive models have been a rapidly growing research area, including data-driven elasticity~\cite{kanno2021kernel}, plasticity~\cite{mozaffar2019deep}, viscoelasticity~\cite{xu2021learning}, etc.
Yet, the current workflow requires using multiple tools and transferring data back and forth, which is cumbersome. 
For instance, simulation data is first generated with certain FEM software, models are then trained using a machine learning library, and finally the trained model is implemented back into the FEM software, often in a hard-coded way.
Built on \texttt{JAX} and having access to all its machine learning functionalities, \texttt{JAX-FEM} provides an ideal platform to solve the sub-problems all in the same ecosystem with high efficiency. \\

The paper is organized as follows. Section \ref{Sec:forward} solves several representative forward problems and the computational performance of \texttt{JAX-FEM} is compared to an open-source FEM software \texttt{FEniCSx}~\cite{logg2012automated} and a commercial software \texttt{Abaqus}.
Section \ref{Sec:inverse} introduces the formulation of solving inverse problems. 
Two applications are presented: full-field reconstruction from sparse observations and structural topology optimization.
Section \ref{Sec:ml} discusses the role of \texttt{JAX-FEM} as an integrated platform for machine learning-enabled computational mechanics. 
One numerical example of data-driven multi-scale computations of composite material is presented.
The three sections correspond to the three features discussed previously.
We then conclude in Section~\ref{Sec:conclusion} with possible future improvement.

Standard notation is used throughout the paper.
Normal fonts are used for scalars, and boldface fonts for vectors (lower-case) and second-order tensors (upper-case).
All tensor and vector components are written with respect to a fixed Cartesian coordinate system with orthonormal basis $\lbrace \boldsymbol{e}_i\rbrace$.
We denote by $\Ib$ the second-order identity tensor.
The prefixes $tr$ and $det$ indicate the trace and the determinant. The superscript ${\top}$ means the transpose of a second-order tensor.
Let ($\boldsymbol a$, $\boldsymbol b$) be vectors, ($\boldsymbol A$, $\boldsymbol B$) be second-order tensors and $\nabla$ the gradient operator; we define the following: $\boldsymbol a \cdot \boldsymbol b = a_{i} b_{i}$, $ (\boldsymbol A \cdot \boldsymbol a)_{i} = A_{il} a_{l}$, $(\boldsymbol A \cdot \boldsymbol B)_{il} = A_{ip}B_{pl}$, 
$\boldsymbol A: \boldsymbol B = A_{il}B_{il}$, $(\nabla \boldsymbol a)_{il} = \partial_l a_i$, $\nabla \cdot \boldsymbol a = \partial_i a_i$ and $ (\nabla \cdot \boldsymbol A )_{i} = \partial_{l} A_{il}$.
We denote by $H^k(\Omega,\mathbb{R}^\textrm{dim})$ the Sobolev space $W^{k,2}(\Omega,\mathbb{R}^\textrm{dim})$ and the square-integrable function space $L^2=H^0$.
Norm is denoted by $\Vert \square \Vert$ for a function while $|\square|$ for a finite-dimensional vector (Euclidean norm).
Note that boldface font is also used for large matrix/vector assembled by FEM, e.g., a vector of nodal degrees of freedom (DOF) $\Ub$ or a matrix~$\Kb$.
We use $\Kb^*$ to denote the adjoint of $\Kb$.

Our code is continuously being developed and available at \href{https://github.com/tianjuxue/jax-am/tree/main/jax_am/fem}{https://github.com/tianjuxue/jax-am/tree/main/jax\_am/fem}.

\section{Solving forward problems}
\label{Sec:forward}

In this section, we first define the class of problems to solve.
Then we discuss several key features of \texttt{JAX-FEM} that are distinguished from the classic implementation of FEM, including array programming style and the use of automatic differentiation technique.
By solving typical solid mechanics problems of linear elasticity, hyperelasticity, and plasticity with both \texttt{JAX-FEM} and \texttt{FEniCSx} and comparing their results, we ensure the correctness of \texttt{JAX-FEM}.
Finally, we conduct a performance test to show the scalability of \texttt{JAX-FEM} for efficiently solving large-size problems of DOF around 7.7 million.

\subsection{Problem statement: nonlinear FEM}

For illustration purposes, let us consider second-order elliptic partial differential equations with the following form: Find $\ub: \Omega \rightarrow \mathbb{R}^{\textrm{vec}}$ such that
\begin{align} \label{Eq:strong}
    -\nabla \cdot \big(\fb (\nabla \ub) \big) = \bb & \quad \textrm{in}  \, \, \Omega, \nonumber \\
    \ub = \ub_D &  \quad\textrm{on} \, \, \Gamma_D,  \nonumber \\
    \fb(\nabla \ub)  \cdot \nb = \tb  & \quad \textrm{on} \, \, \Gamma_N.
\end{align}
where $\Omega\subset\mathbb{R}^{\textrm{dim}}$ is the problem domain, $\bb$ is the source term, $\ub_D$ is the Dirichlet boundary condition defined on ${\Gamma_D\subset\partial\Omega}$, $\nb$ is the outward normal, $\tb$ prescribes Neumann boundary condition on ${\Gamma_N\subset\partial\Omega}$ (${\Gamma_D\cup\Gamma_N=\partial\Omega}$ and ${\Gamma_D\cap\Gamma_N=\emptyset}$), and $\fb: \mathbb{R}^{\textrm{vec}\times \textrm{dim}} \rightarrow \mathbb{R}^{\textrm{vec}\times \textrm{dim}}$ is a general tensor-valued function that governs the physics of the problem.
Here, ``vec'' is the number of vector variable components, and ``dim'' is the spatial dimension.
In this work, we only consider three-dimensional problems with ${\textrm{dim}=3}$.

The weak form of Eq.~(\ref{Eq:strong}) reads the following: Find $\ub\in\mathcal{U}$ such that $\forall \vb \in \mathcal{V}$
\begin{align} \label{Eq:weak}
  F(\ub; \vb) = \int_{\Omega}  \fb(\nabla \ub) :  \nabla \vb \textrm{ d} \Omega - \int_{\Gamma_N} \tb \cdot  \vb \textrm{ d} \Gamma - \int_{\Omega} \bb \cdot  \vb \textrm{ d} \Omega = 0,
\end{align}
where the trial and test function spaces are 
\begin{align}
    \mathcal{U} &= \big\lbrace \ub \in H^1(\Omega,\, \mathbb{R}^{\textrm{dim}}) \,\, \big| \,\, \ub = \ub_D \textrm{ on } \Gamma_D \big\rbrace,  \nonumber \\
    \mathcal{V} &= \big\lbrace \vb \in H^1(\Omega,\, \mathbb{R}^{\textrm{dim}}) \,\, \big| \,\, \vb = \textbf{0} \textrm{ on } \Gamma_D \big\rbrace.
\end{align}

Newton's method for solving the nonlinear problem (\ref{Eq:weak}) yields the following linearized incremental problem: Find the incremental solution $\delta\ub\in\mathcal{V}$ such that
\begin{align} \label{Eq:newton}
    F'(\ub, \delta \ub; \vb) = -F(\ub; \vb),
\end{align}
where the Gateaux derivative is defined as
\begin{align} \label{Eq:linearization}
    F'(\ub, \delta \ub; \vb) = \lim_{\epsilon \rightarrow 0} \frac{F(\ub + \epsilon \delta \ub; \vb) - F(\ub; \vb)}{\epsilon} = \int_{\Omega} \nabla \vb : \mathbb{C} : \nabla\delta \ub \textrm{ d} \Omega,
\end{align}
where $\mathbb{C}$ is the fourth-order tangent tensor with  $C_{ijkl} = f_{ij,kl}$.
The Galerkin finite element method discretizes Eq.~(\ref{Eq:newton}) so that the following finite-dimensional linear system is solved:
\begin{align} \label{Eq:galerkin}
    F'(\ub^h, \delta \ub^h; \vb^h) = -F(\ub^h; \vb^h),
\end{align}
where $u^h \in \mathcal{U}^h \subset \mathcal{U}$ is the current solution, $\delta u^h \in \mathcal{V}^h \subset \mathcal{V}$ is the incremental solution we need to solve for, and $v^h \in \mathcal{V}^h$ is the test function.
Here, $\mathcal{U}^h$ and $\mathcal{V}^h$ are the finite element function spaces.

\subsection{Array programming and automatic differentiation}

Central to FEM implementation is to assemble the linear system corresponding to Eq.~(\ref{Eq:galerkin}).
The procedure is shown in Alg.~\ref{Alg:assembly}, where $N_e$ is the total number of elements and $N_d$ is the number of DOF associated with each element.

\begin{algorithm}[H]
\caption{Conceptual process of matrix assembly}\label{Alg:assembly}
Initialize global stiffness matrix $\Kb$ \tcp{Sparse matrix} 
\For{$e\gets1$  \KwTo $N_e$}{
    Initialize element stiffness matrix $\Kb_e$ \tcp{Dense matrix of size $N_d\times N_d$} 
    \For{$i\gets1$ \KwTo $N_d$}
    {
        \For{$j\gets1$ \KwTo $N_d$} 
        {
            Update $\Kb_e[i, j]$ 
        }
    }
    Add $\Kb_e$ to $\Kb$
}
Return $\Kb$
\end{algorithm}
\vspace{10pt}

When implementing Alg.~\ref{Alg:assembly}, there are two features that fundamentally distinguish \texttt{JAX-FEM} from traditional approaches: array programming and automatic differentiation. 

\paragraph{Array programming}

While for-loops exist in the conceptual illustration of Alg.~\ref{Alg:assembly}, we never explicitly write any of these for-loops as common practice in \texttt{Fortran} or \texttt{C/C++} implementation.
Instead, array programming style~\cite{van2011numpy} (in the same spirit of \texttt{NumPy}~\cite{harris2020array}) is used for fully utilizing the power of GPU acceleration.
The implementation makes use of \texttt{jax.vmap}, a core function of \texttt{JAX} for vectorized operations.

\paragraph{Automatic differentiation}

Classic FEM implementation requires computing explicitly the entries of the element stiffness matrix $\Kb_e$ such that
\begin{align}
    \Kb_e[i, j] = \int_{\Omega_e} \nabla \phib_i : \mathbb{C}: \nabla \phib_j \textrm{ d} \Omega,
\end{align}
where $\phib_i$ is the $i$th FEM basis function restricted to the element domain $\Omega_e$.
\texttt{JAX-FEM} does not require explicitly deriving the linearized form as in Eq.~(\ref{Eq:linearization}).
Instead, the program works directly with the weak form defined in Eq.~(\ref{Eq:weak}). 
We define the element residual vector function $\rb_e$ as
\begin{align}
    \rb_e: \mathbb{R}^{N_d} &\rightarrow \mathbb{R}^{N_d}, \nonumber \\
     \Ub_e &\mapsto \Rb_e, \nonumber \\
     \Rb_e[i] & = \int_{\Omega_e}  \fb(\nabla \ub^h) :  \nabla \phib_i \textrm{ d} \Omega,
\end{align}
where $\Ub_e$ is the vector of nodal DOF defined in the element, $\Rb_e$ is the residual vector, and $\ub^h(\x) = \sum_k^{N_d} \Ub_e[k] \phib_k(\x)$ is the FEM solution field.
To obtain $\Kb_e[i, j]$, we simply compute the Jacobian matrix of $\rb_e$ at $\Ub_e$ such that
\begin{align}
     \Kb_e[i, j] = \frac{\partial \rb_e}{\partial \Ub_e}[i, j]. 
\end{align}
We use automatic differentiation provided by \texttt{JAX} to compute this Jacobian matrix. 
In many applications, e.g., plasticity, the fourth-order tangent tensor $\mathbb{C}$ is nontrivial to derive, and \texttt{JAX-FEM} frees developers from this tedious procedure.

\subsection{Linear elasticity, hyperelasticity, and plasticity}

\begin{figure}[H] 
\centering
\includegraphics[scale=0.8]{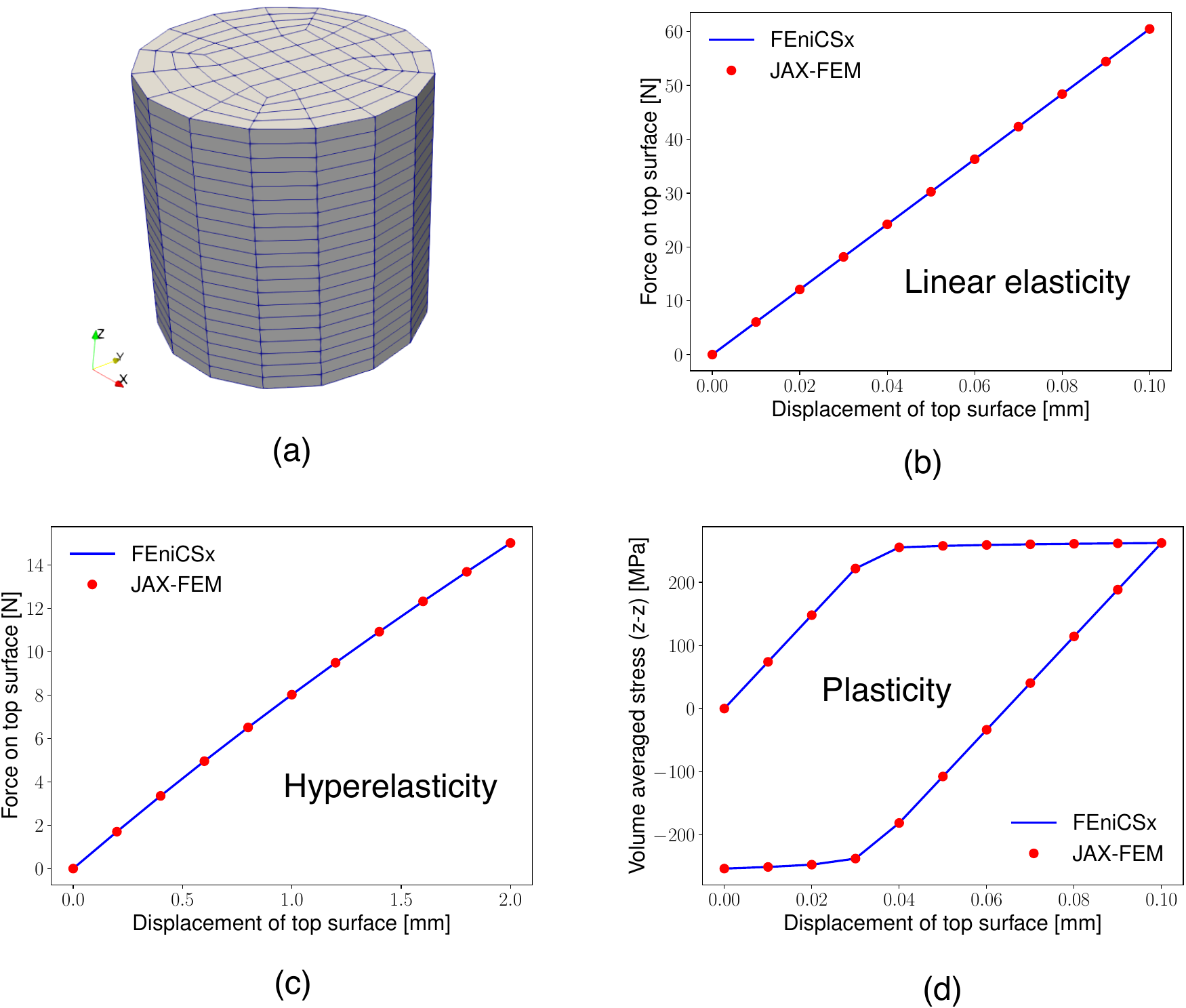}
\caption{Testing \texttt{JAX-FEM} with \texttt{FEniCSx} being the ground truth, where (a) is the problem mesh, (b) considers a linear elastic material, (c) considers a hyperelastic material, and (d) considers an elasto-plastic material.}
\label{Fig:forward}
\end{figure}

To verify the correctness of \texttt{JAX-FEM}, we consider three typical solid mechanics problems, i.e., linear elasticity, hyperelasticity, and plasticity.
Specifically, we impose uniaxial tensile loadings on a cylinder (see Fig.~\ref{Fig:forward} (a)) where the bottom boundary is fixed and the top boundary is subject to fixed displacement conditions.
The height of the cylinder is 10\,mm and the radius is 5\,mm.
For simplicity, we assume zero body force and free traction force for all three problems.
We solve the problems with both \texttt{JAX-FEM} and \texttt{FEniCSx} and compare the results.

\paragraph{Linear elasticity} 
We replace the tensor function $\fb$ in Eq.~(\ref{Eq:strong}) with the Cauchy stress $\cs$ so that the governing equation is
\begin{align} \label{Eq:hyperelastic_strong}
    -\nabla \cdot \cs = \boldsymbol{0} & \quad \textrm{in}  \, \, \Omega, \nonumber \\
    \ub = \ub_D &  \quad\textrm{on} \, \, \Gamma_D,  \nonumber \\
    \cs \cdot \nb = \boldsymbol{0}  & \quad \textrm{on} \, \, \Gamma_N,
\end{align}
where we have 
\begin{align} \label{Eq:linear_elasticity}
     \cs &=  \lambda \, \textrm{tr}(\vepsb) \Ib + 2\mu \, \vepsb, \nonumber \\
    \vepsb &= \frac{1}{2}\left[\nabla\ub + (\nabla\ub)^{\top}\right], \nonumber \\
       \fb(\nabla \ub) &= \cs,
\end{align}
where $\boldsymbol{I}$ is the identity tensor and $\lambda$ and $\mu$ are the Lam\'e parameters.
We assume quasi-static incremental loadings from 0 to 0.1\,mm with 10 steps, and show the plot of force versus displacement in Fig.~\ref{Fig:forward} (b), where the results agree well between \texttt{JAX-FEM} and \texttt{FEniCSx}.

\paragraph{Hyperelasticity}
For a typical neo-Hookean solid, the governing equation is 
\begin{align} \label{Eq:hyperelastic_strong}
    -\nabla \cdot \Pb = \boldsymbol{0} & \quad \textrm{in}  \, \, \Omega, \nonumber \\
    \ub = \ub_D &  \quad\textrm{on} \, \, \Gamma_D,  \nonumber \\
    \Pb \cdot \nb = \boldsymbol{0}  & \quad \textrm{on} \, \, \Gamma_N,
\end{align}
where $\Pb$ is the first Piola-Kirchhoff stress.
Eq.~(\ref{Eq:hyperelastic_strong}) is simply a specific form of Eq.~(\ref{Eq:strong}) in the sense that $\fb$ can be defined through
\begin{align} \label{Eq:neo-hookean}
    \Pb &= \frac{\partial W}{\partial \Fb}, \nonumber \\
    \Fb &= \nabla \ub + \Ib, \nonumber \\
    W (\Fb) &= \frac{G}{2}(J^{-2/3} I_1 - 3) + \frac{\kappa}{2}(J - 1)^2, \nonumber \\
    \fb(\nabla \ub) &= \Pb,
\end{align}
where $\Fb$ is the deformation gradient, $W$ is the strain energy density function, $J = \textrm{det}(\Fb)$,~$I_1 = \textrm{tr}(\boldsymbol{C})$;~$G = \frac{E}{2(1+\nu)}$ and~$\kappa = \frac{E}{3(1-2\nu)}$ denote the initial shear and bulk moduli, respectively, with~$E$ being the Young's modulus and and~$\nu$ the Poisson's ratio of the material.
The above~${W}$ is commonly used to model isotropic elastomers that are almost incompressible~\cite{ogden1997non}.
We assume quasi-static incremental loadings from 0 to 2\,mm with 10 steps.
We show the plot of force versus displacement in Fig.~\ref{Fig:forward} (c), where the results agree well between \texttt{JAX-FEM} and \texttt{FEniCSx}. 

\paragraph{Plasticity} 
For perfect J2-plasticity model~\cite{simo2006computational}, we assume that the total strain $\vepsb^{k-1}$ and stress $\cs^{k-1}$ from the previous loading step are known, and the problem states that find the displacement field $\ub^k$ at the current loading step such that
\begin{align} \label{Eq:plasticity_strong}
    -\nabla \cdot \big(\fb (\nabla \ub^k, \vepsb^{k-1}, \cs^{k-1}) \big) = \boldsymbol{0} & \quad \textrm{in}  \, \, \Omega, \nonumber \\
    \ub^k = \ub_D &  \quad\textrm{on} \, \, \Gamma_D,  \nonumber \\
    \fb \cdot \nb = \boldsymbol{0}  & \quad \textrm{on} \, \, \Gamma_N.
\end{align}
Eq.~(\ref{Eq:plasticity_strong}) is a specific form of Eq.~(\ref{Eq:strong}), where the function $\fb$ is defined with the following plasticity-related equations
\begin{align}
    \cs_\textrm{trial} &= \cs^{k-1} + \Delta \cs, \nonumber\\
    \Delta \cs &= \lambda \, \textrm{tr}(\Delta \vepsb) \Ib + 2\mu \, \Delta \vepsb, \nonumber \\
    \Delta \vepsb &= \vepsb^k  - \vepsb^{k-1} = \frac{1}{2}\left[\nabla\ub^k + (\nabla\ub^k)^{\top}\right] - \vepsb^{k-1}, \nonumber\\
    \sbf &= \cs_\textrm{trial} - \frac{1}{3}\textrm{tr}(\cs_\textrm{trial})\Ib,\nonumber\\
    s &= \sqrt{\frac{3}{2}\sbf:\sbf}, \nonumber\\
    f_{\textrm{yield}} &= s - \sigma_{\textrm{yield}}, \nonumber\\
    \cs^k &= \cs_\textrm{trial} -  \frac{\sbf}{s} \langle f_{\textrm{yield}} \rangle_{+}, \nonumber \\
    \fb (\nabla \ub^k, \vepsb^{k-1}, \cs^{k-1}) &= \cs^k,
\end{align}
where $\cs_\textrm{trial}$ is the elastic trial stress, $\sbf$ is the devitoric part of $\cs_\textrm{trial}$, $f_{\textrm{yield}}$ is the yield function, $\sigma_{\textrm{yeild}}$ is the yield strength, ${\langle x \rangle_{+}}:=\frac{1}{2}(x+|x|)$ is the ramp function, and $\cs^k$ is the stress at the currently loading step.
Deriving the four-order elastoplastic tangent moduli tensor $\mathbb{C}$ is usually required by traditional FEM implementation, but is not needed by \texttt{JAX-FEM} due to automatic differentiation.
We assume quasi-static loadings from 0 to 0.1\,mm and then unload from 0.1\,mm to 0.
We show the plot of the $z$-$z$ component of volume-averaged stress versus displacement of the top surface in Fig.~\ref{Fig:forward} (d), where the path-dependent results match exactly between \texttt{JAX-FEM} and \texttt{FEniCSx}.

Besides the comparison results above, we have maintained a    \href{https://github.com/tianjuxue/jax-am/tree/main/jax_am/fem/tests}{test suite} in our open source repository for more tests like body force, Neumann boundary conditions, etc., so that each new version of \texttt{JAX-FEM} must pass these tests.

\subsection{A sample user program}

The interface of \texttt{JAX-FEM} for users is succinct.
Here, we revisit the tensile problem for the cylinder with linear elastic material.
As shown in the code snippet below, users first create the cylinder mesh.
Dirichlet boundary conditions are imposed component-wisely, i.e., we need to specify conditions for bottom-$x$, bottom-$y$, bottom-$z$, top-$x$, top-$y$, and top-$z$ separately.
Then, \texttt{mesh} and \texttt{dirichlet\_bc\_info} are passed to the FEM model \texttt{LinearElasticity} and the \texttt{solver} solves the problem

\begin{python}
import jax.numpy as np
from jax_am.fem.models import LinearElasticity
from jax_am.fem.solver import solver
from jax_am.fem.generate_mesh import cylinder_mesh 

mesh = cylinder_mesh()

bottom = lambda point: np.isclose(point[2], 0.)
top = lambda point: np.isclose(point[2], 10.)
zero_disp = lambda point: 0.
top_z_disp = lambda point: 0.1

dirichlet_bc_info = [[bottom, bottom, bottom, top, top, top], 
                     [0, 1, 2, 0, 1, 2], 
                     [zero_disp, zero_disp, zero_disp, 
                      zero_disp, zero_disp, top_z_disp]]

problem = LinearElasticity(mesh, dirichlet_bc_info)
solution = solver(problem)
\end{python}

\subsection{Performance and scalability}

\begin{figure}[H] 
\centering
\includegraphics[scale=0.5]{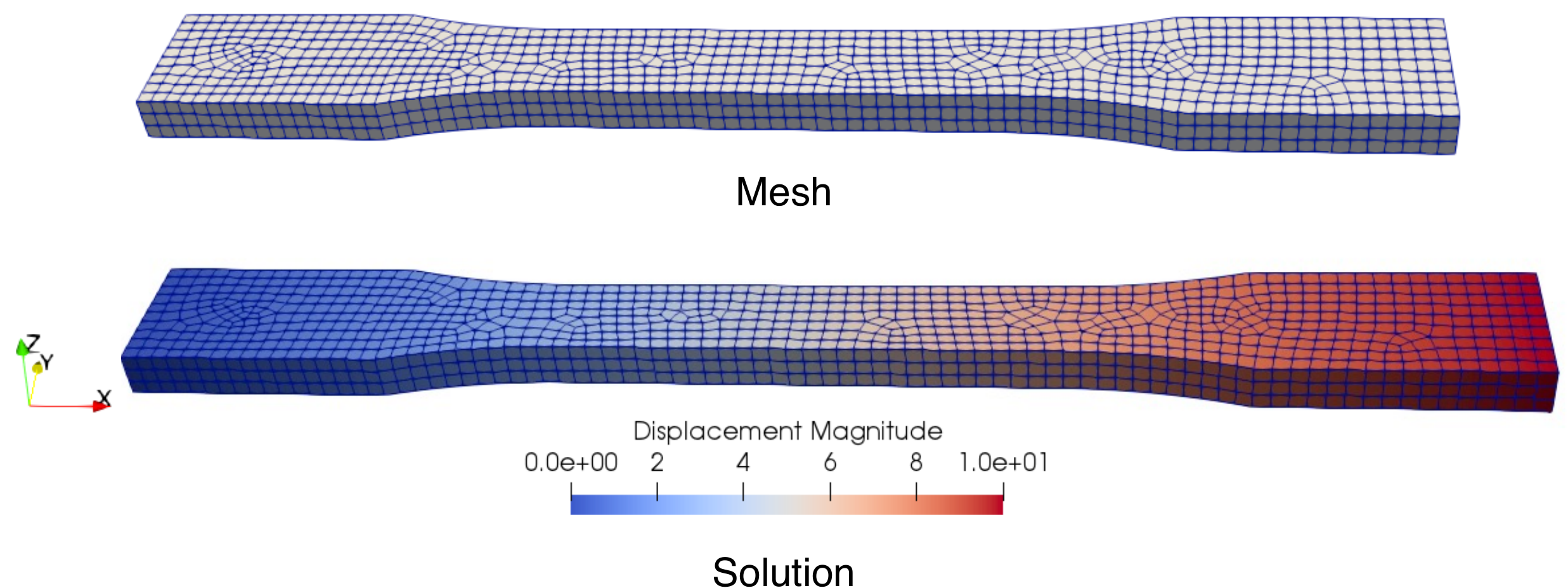}
\caption{The dog bone shaped ASTM D638 Type 1 specimen, which is a standard test specimen for tensile properties of polymers~\cite{Gooch2011}. The tensile experiment fixes the left side and pulls the right side with a prescribed displacement condition.}
\label{Fig:dogbone}
\end{figure}

\texttt{JAX-FEM} is based on \texttt{JAX} and uses just-in-time compilation for high performance.
Therefore, the fact that \texttt{JAX-FEM} is written in pure \texttt{Python} does not limit us to only small problems.
For benchmarking the performance, we solve a standard uniaxial tensile loading problem on an ASTM D638 Type 1 specimen (see Fig.~\ref{Fig:dogbone}) assuming linear elastic material.

\begin{figure}[H] 
\centering
\includegraphics[scale=0.4]{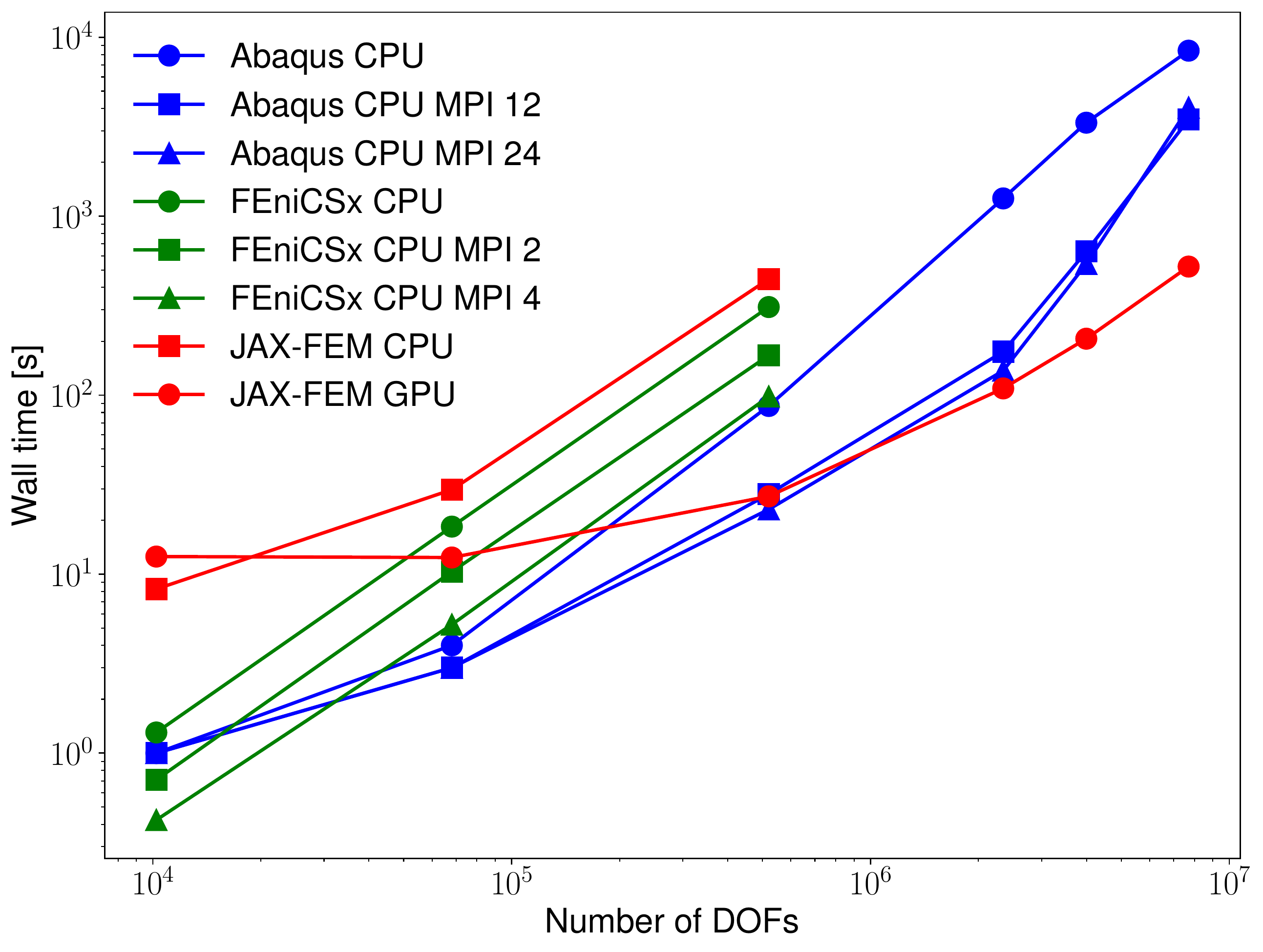}
\caption{Performance report. Here, ``FEniCSx CPU MPI 4'' means \texttt{FEniCSx} runs with 4 processes of MPI parallel programming.}
\label{Fig:performance}
\end{figure}

The same problem with different levels of mesh resolution is solved using \texttt{JAX-FEM} with CPU-only mode and with GPU, \texttt{FEniCSx} with MPI for parallel programming, and \texttt{Abaqus} running on CPU with/without MPI.
The wall time measurements with respect to the number of DOF are shown in Fig.~\ref{Fig:performance}.
\texttt{JAX-FEM} shows a predominant advantage when GPU is used.
The largest problem has 7,703,841 DOF and takes 8409\,s and 4769\,s for \texttt{Abaqus} with CPU and MPI (24 cores), respectively, and 523\,s for \texttt{JAX-FEM} on GPU. \texttt{JAX-FEM} on GPU achieves 16.1$\times$ and 9.1$\times$ acceleration compared to \texttt{Abaqus} on CPU and with MPI, respectively. The problem in Fig.~\ref{Fig:dogbone} has 10,224 DOF and corresponds to the first column of data points in Fig.~\ref{Fig:performance}.

Note that for \texttt{Abaqus} the MPI acceleration is not significant as the number of DOF becomes larger. For example, the \texttt{Abaqus} MPI speedup compared with \texttt{Abaqus} CPU is 9.1$\times$ for 2,344,230 DOF but 1.8$\times$ for 7,703,841 DOF. 
This decreased performance in Abaqus as DOF increases is attributed to the severe message passing delay for large DOF problems. In \texttt{Abaqus}, if the size of transient variables (mainly the global stiffness matrix) exceeds CPU memory limit, they are stored on local storage where most of the delay takes place. 
It is also worth mentioning that with 24 cores, we need 19 paid tokens. Compared to this, JAX-FEM is an open-source software and faster than \texttt{Abaqus} with an extended license for tokens.

We report the platforms for those numerical experiments.
\texttt{JAX-FEM} runs on 2.3 GHz Intel(R) Xeon(R) W-2195 CPU (18 cores) with NVIDIA Quadro RTX 8000 GPU (48 GB Graphics memory) on Ubuntu 20.04.5 LTS.
\texttt{FEniCSx} runs on 2.4 GHz Intel i9 CPU (8 cores) on macOS Big Sur 11.6.5.
\texttt{Abaqus} CPU runs on Intel(R) Core(TM) i7-4790 CPU @ 3.60 GHz under Windows operating system. 
\texttt{Abaqus} CPU with MPI runs on Intel(R) Xeon (R) CPU E5-2680 v3 @ 2.50 GHz (total 24 cores) with 19 tokens under CentOS 6.9 operating system.

\section{Solving inverse problems}
\label{Sec:inverse}

We formulate inverse problems as PDE-constrained optimization (PDE-CO) problems.
There are in general two strategies for solving PDE-CO problems: optimize-then-discretize and discretize-then-optimize~\cite{betts2005discretize,liu2019non}.
We follow the discretize-then-optimize approach.
The discretized PDE-CO problem is formulated as
\begin{align}
    \label{Eq:PDE_constraint}
    \nonumber \min_{\Ub\in\sR^{N}, \thetab\in\sR^{M}} \gJ(\Ub, \thetab) \\
    \textrm{s.t.} \quad \Cb(\Ub, \thetab)=\boldsymbol{0}, 
\end{align}
where $\Ub$ is the finite element solution vector of DOF, $\thetab$ is the parameter vector, and ~$\gJ(\cdot, \cdot): \sR^{N}\times \sR^{M} \rightarrow \sR$ is the objective function.
The constraint function $\Cb(\cdot, \cdot): \sR^{N}\times \sR^{M} \rightarrow \sR^{N}$ represents the discretized governing PDE and should be regarded as the direct consequence of discretizing the weak form in Eq.~(\ref{Eq:weak}) and imposing Dirichlet boundary conditions.

A reduced formulation is used to embed the PDE constraint so that the problem posed in (\ref{Eq:PDE_constraint}) is re-formulated as
\begin{align}
    \label{Eq:adjoint_obj}
    \min_{\thetab\in\sR^{M}}\widehat{\gJ}(\thetab),
\end{align}
where~${\widehat{\gJ}(\thetab):= \gJ(\Ub(\thetab), \thetab)}$ and~${\Ub(\thetab)}$ is the implicit function that arises from solving the PDE.
For efficient optimization algorithms, gradient information is necessary.
The total derivative of $\widehat{\gJ}$ with respect to parameters $\thetab$ is computed with chain rules
\begin{align}
    \label{Eq:chain}
    \frac{\textrm{d} \widehat{\gJ}}{\textrm{d} \thetab} = \frac{\partial \gJ}{\partial \Ub}\frac{\textrm{d} \Ub}{\textrm{d} \thetab} +\frac{\partial \gJ}{\partial \thetab},
\end{align}
where the existence of the derivative $\frac{\textrm{d} \Ub}{\textrm{d} \thetab}$ is justified by the implicit function theorem~\cite{MR0055409} under certain mild conditions.
We then take the derivative of the constraint function in Eq.~(\ref{Eq:PDE_constraint}) with respect to $\thetab$ so that the following relations are obtained
\begin{align}
    \dfrac{\textrm{d} \Cb}{\textrm{d} \thetab} = \dfrac{\partial \Cb}{\partial \Ub} \dfrac{\textrm{d} \Ub}{\textrm{d} \thetab} + \dfrac{\partial \Cb}{\partial \thetab}=0.
\end{align}
Hence,
\begin{align}
    \label{Eq:implicit}
    \dfrac{\textrm{d} \Ub}{\textrm{d} \thetab} = - \Big(\dfrac{\partial \Cb}{\partial \Ub}\Big)^{-1} \dfrac{\partial \Cb}{\partial \thetab}.
\end{align}
Substitute Eq.~(\ref{Eq:implicit}) to Eq.~(\ref{Eq:chain}), we obtain
\begin{align}
    \label{Eq:adjoint_chain}
    \frac{\textrm{d} \widehat{\gJ}}{\textrm{d} \thetab} = -  \rlap{$\overbrace{\phantom{\frac{\partial \gJ}{\partial \Ub} \Big(\dfrac{\partial \Cb}{\partial \Ub}\Big)^{-1}}}^{\textrm{adjoint PDE}}$} \frac{\partial \gJ}{\partial \Ub}  \underbrace{\Big(\dfrac{\partial \Cb}{\partial \Ub}\Big)^{-1} \dfrac{\partial \Cb}{\partial \thetab}}_{\textrm{tangent linear PDE}}  +\frac{\partial \gJ}{\partial \thetab},   
\end{align}
where the first term can be evaluated either from left to right (solving the adjoint PDE first) or from right to left (solving the tangent linear PDE first).
When the size of the parameter vector $\thetab$ is larger than that of the objective (e.g.,~${M \gg 1}$ in our case), it is more efficient to solve the adjoint PDE first, giving the name adjoint method~\cite{cao2003adjoint}.
For interested readers, Xu and Darve~\cite{xu2022physics} recently presented a detailed discussion on the cost comparison of the adjoint method and the tangent linear approach.
We continue the discussion by adopting the adjoint method.
The adjoint PDE is 
\begin{align} \label{Eq:adjoint_PDE}
    \frac{\partial \Cb}{\partial \Ub}^* \lambdab = \frac{\partial \mathcal{J}}{\partial \Ub}^*,
\end{align}
where $\lambdab \in \mathbb{R}^{N}$ is the adjoint variable.
Substitute $\lambdab$ to Eq.~(\ref{Eq:adjoint_chain}) we have 
\begin{align}
    \label{Eq:adjoint_chain_simp}
    \frac{\textrm{d} \widehat{\gJ}}{\textrm{d} \thetab} = -\lambdab^* \dfrac{\partial \Cb}{\partial \thetab} +\frac{\partial \gJ}{\partial \thetab}.
\end{align}
Note that Eq.~(\ref{Eq:adjoint_PDE}) is a linear PDE to solve, but it relies on the Jacobian matrix $\frac{\partial \Cb}{\partial \Ub}$, which requires the solution vector $\Ub$. 
Therefore, the computational cost is largely dominated by solving the forward problem, not the adjoint PDE, especially when the forward PDE is nonlinear.

The derivations above are abstract and problem independent.
When solving specific problems, one typically needs to further derive the concrete expressions of those derivatives, e.g., $\frac{\partial \Cb}{\partial \thetab}$, which is often tedious and error-prone.
In \texttt{JAX-FEM}, we use automatic differentiation to compute these derivatives, which greatly enhances productivity.
The following code snippet demonstrates an example of given $\Cb(\cdot, \thetab)$ (\texttt{partial\_constraint\_fn}) and computing $\lambdab^*\frac{\partial \Cb}{\partial \thetab}$ (\texttt{result}) using \texttt{JAX} function \texttt{jax.vjp}, which stands for vector-Jacobian product.
\begin{python}
# params: JAX array of shape (M,)
# adjoint: JAX array of shape (N,)
def vec_jac_prod_fn(v):
    primals, vec_jac_prod = jax.vjp(partial_constraint_fn, params)
    val, = vec_jac_prod(v)
    return val
result = vec_jac_prod_fn(adjoint) 
\end{python}
Similar approaches are applied to other derivative-related computations like $\frac{\partial \Cb}{\partial \Ub}^* \lambdab$ in Eq.~(\ref{Eq:adjoint_PDE}) so that the entire workflow of computing the total derivative $\frac{\textrm{d} \widehat{\gJ}}{\textrm{d} \thetab}$ is fully automatic.
For general information about automatic differentiation involving implicit functions, we refer to the recent work by Blondel et al.~\cite{blondel2021efficient}.
As a summary, we show the overall workflow of solving PDE-CO problems in Alg.~\ref{Alg:pde_co}.
In general, the ``optimizer'' in the algorithm can use any off-the-shelf gradient-based optimization algorithms.
The simplest one is the gradient descent method, but more sophisticated algorithms are usually required given the complexity of the specific problem.

\begin{algorithm}[H]
\caption{PDE-constrained optimization with the adjoint method}\label{Alg:pde_co}
\KwInput{$\thetab_{\textrm{ini}}$, $i_{\rm{max}}$ \tcp{Initial parameter and maximum iteration number} } 
$\thetab\leftarrow \thetab_{\rm{ini}}$, $i\leftarrow 0$ \\
\While{$i < i_{\rm{max}}$}{
$\Ub$ $\leftarrow$ ForwardPDESolver($\thetab$)  \tcp{See constraint function in Eq.~(\ref{Eq:PDE_constraint})}
$\frac{\textrm{d} \widehat{\gJ}}{\textrm{d} \thetab}$ $\leftarrow$ AdjointMethod($\Ub$, $\thetab$) \tcp{See Eq.~(\ref{Eq:adjoint_obj}) to Eq.~(\ref{Eq:adjoint_chain_simp})}
$\thetab$ $\leftarrow$ Optimizer($\thetab$, $\frac{\textrm{d} \widehat{\gJ}}{\textrm{d} \thetab}$) \tcp{Gradient-based optimizer}
$i \leftarrow i+1$
}
\KwOutput{$\Ub$, $\thetab$}
\end{algorithm}

In the next two subsections, we pose specific PDE-CO problems in the form of Eq.~(\ref{Eq:PDE_constraint}) and use \texttt{JAX-FEM} for solutions.


\subsection{Full field inference from sparse observations}

In this numerical example, we consider predicting the full scalar field with sparse observations at certain randomly picked points.
The forward governing PDE is a linear Poisson's equation:
\begin{align} \label{Eq:full_field_forward}
    -\alpha \Delta u = b & \quad \textrm{in}  \, \, \Omega, \nonumber \\
    u = 0 &  \quad\textrm{on} \, \, \partial \Omega,
\end{align}
where $\alpha$ is a constant coefficient and $b$ is the source function that we can control.
The weak form states that find the solution $u$ so that for any test function $v$ we have
\begin{align} \label{Eq:full_field_weak}
\int_{\Omega} \alpha \nabla u \cdot \nabla v \textrm{ d} \Omega  = \int_{\Omega} b \, v \textrm{ d} \Omega,
\end{align}
The PDE-CO problem states that 
\begin{align} \label{Eq:full_field_inverse}
    \nonumber \min_{\Ub\in\sR^{N}, \thetab\in\sR^{M}} \sum_{i\in \mathcal{I}_{\textrm{obs}}} (\Ub[i] - U_{i, \textrm{obs}})^2 \\
    \textrm{s.t.} \quad \Ab \, \Ub = \Fb (\thetab), 
\end{align}
where $\Ub$ is the DOF vector of $u$, $\thetab$ is the discretized version of $b$, $\Ub[i]$ is the $i$th component of $\Ub$, $U_{i,\textrm{obs}}$ is the $i$th observed value, $\mathcal{I}_{\textrm{obs}}$ is the index set of observations, and the constraint equation $\Ab \, \Ub = \Fb(\thetab)$ is the discretized version of the weak form (\ref{Eq:full_field_weak}).

\begin{figure}[H] 
\centering
\includegraphics[scale=0.65]{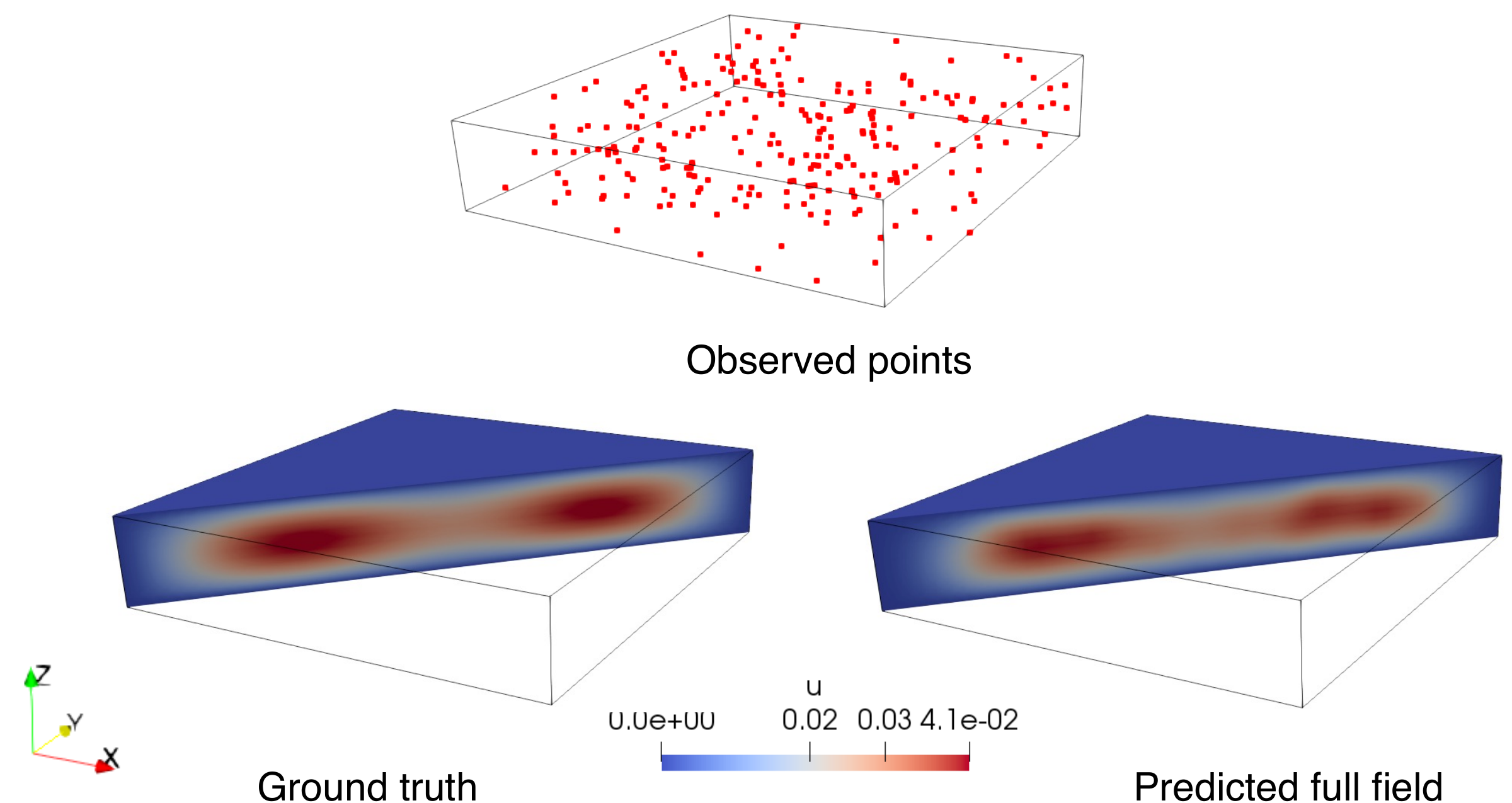}
\caption{Configurations and results of the full field prediction example. The top panel shows the 250 points where the solution value can be observed. The left panel shows the ground truth solution. The right panel shows the predicted solution field with PDE-CO.}
\label{Fig:poisson_setup}
\end{figure}

The problem domain $\Omega$ is a $1\times1\times0.2$ rectangular box discretized with $50\times50\times10$ linear hexahedral elements.
We randomly pick 250 points for observing the true solution values, as shown in the top panel of Fig.~\ref{Fig:poisson_setup}.
The ground truth solution obtained from solving Eq.~(\ref{Eq:full_field_forward}) is shown in the lower left panel of Fig.~\ref{Fig:poisson_setup}.
The source term function $b$ is set to have a bimodal shape such that $b(\x)=10\cdot\textrm{exp}(-10\cdot|\x - (2.5, 2.5, 5.0)|^2) + 10\cdot\textrm{exp}(-10\cdot|\x - (7.5, 7.5, 5.0)|^2)$.
The predicted solution by solving the PDE-CO problem is shown in the lower right panel Fig.~\ref{Fig:poisson_setup}.

\begin{figure}[H] 
\centering
\includegraphics[scale=0.65]{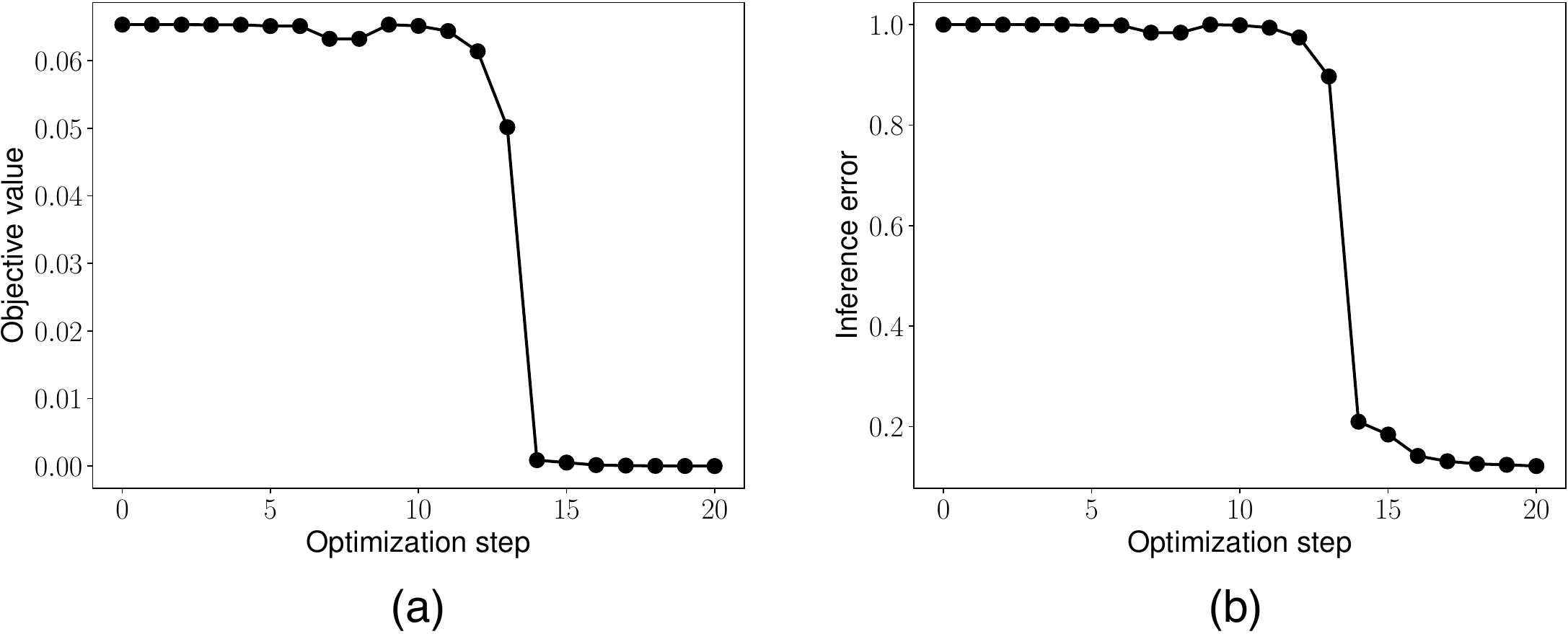}
\caption{Optimization results of the full field prediction example. Subfigure (a) shows the optimization objective converging to zero, and subfigure (b) shows the relative inference error of the predicted solution to the true solution. }
\label{Fig:poisson_opt}
\end{figure}

We show the optimization iterations of the PDE-CO problem in Fig.~\ref{Fig:poisson_opt} (a).
In the plot, the objective value (y-axis) is defined in Eq.~(\ref{Eq:full_field_inverse}) and the optimization step (x-axis) is defined as each time the gradient information is queried by the optimizer (see Alg.~\ref{Alg:pde_co}).
In this problem, the limited-memory BFGS algorithm~\cite{byrd1995limited} provided by the \texttt{SciPy}~\cite{virtanen2020scipy} package is used as the optimizer.
As shown, the objective value quickly drops to nearly zero within only 20 steps.
To quantitatively show the error of the predicted full field solution $u_{\textrm{pred}}$ compared with the ground truth $u_{\textrm{true}}$, we define the relative $L^2$ norm error $\frac{\Vert u_{\textrm{pred}} - u_{\textrm{true}} \Vert_{L^2} }{\Vert u_{\textrm{true}} \Vert_{L^2}}$ and show this inference error in Fig.~\ref{Fig:poisson_opt} (b).
As seen, the error is about 12.0$\%$ when the optimization is over.
Note that this is the result of observing only 250 points, which is less than 1$\%$ of the total points, and we anticipate the error to decrease if more points are observed.
For example, with 2500 observed points, the error is decreased to about 1.4$\%$.

\begin{figure}[H] 
\centering
\includegraphics[scale=0.45]{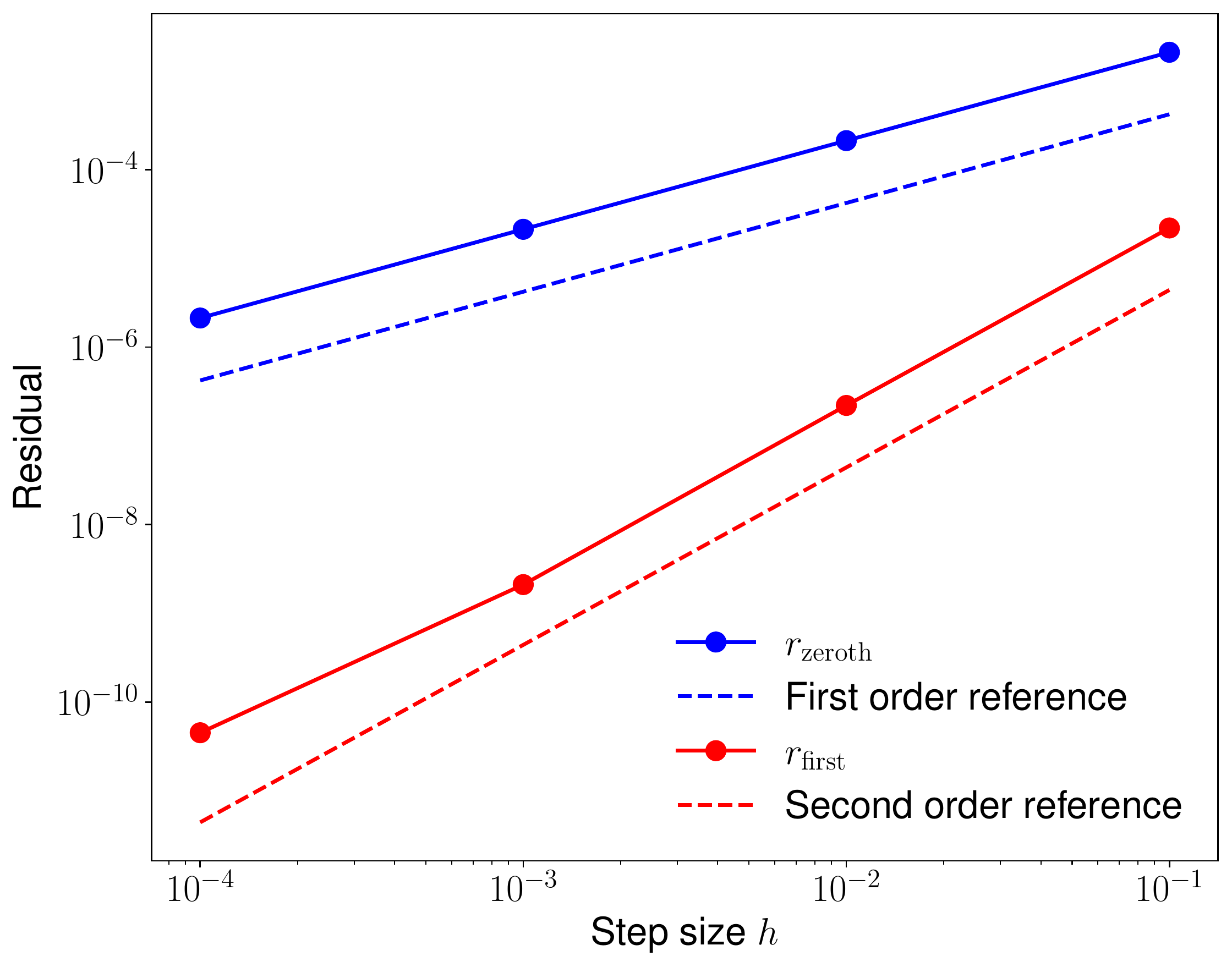}
\caption{Taylor test results. As expected, the zeroth-order expansion of the residual achieves a first order convergence, and the first-order expansion achieves a second order convergence.}
\label{Fig:poisson_res}
\end{figure}

In this PDE-CO problem, the total derivative $\frac{\textrm{d} \widehat{\gJ}}{\textrm{d} \thetab}$ is computed automatically by \texttt{JAX-FEM}.
We perform a sanity test here to ensure that this derivative is computed correctly.
Following the works of~\cite{niewiarowski2020adjoint,xue2022mapped}, Taylor test can be used to check the accuracy of the computed gradient.
Given a perturbation $\delta \thetab$, the convergence rate of the residual should be 1 using a zeroth-order expansion:
\begin{align} \label{Eq:zeroth_expansion}
    r_{\textrm{zeroth}}=\big| \widehat{J}(\thetab + h\delta\thetab) - \widehat{J}(\thetab) \big| \rightarrow 0 \textrm{ at } \mathcal{O}(h),
\end{align}
and the convergence rate should be 2 by a first-order expansion:
\begin{align}\label{Eq:first_expansion}
    r_{\textrm{first}}=\big| \widehat{J}(\thetab + h\delta\thetab) - \widehat{J}(\thetab) - h \frac{\textrm{d}\widehat{J}}{\textrm{d}\thetab}\cdot\delta \thetab \big| \rightarrow 0 \textrm{ at } \mathcal{O}(h^2).
\end{align}
The results above are the direct consequence of Taylor's theorem~\cite{MR0055409}.
We set the step size $h$ to be $h=10^{-4}, 10^{-3}, 10^{-2}, 10^{-1}$ and calculate $r_{\textrm{zeroth}}$ and $r_{\textrm{first}}$, and show the results in Fig.~\ref{Fig:poisson_res}.
As expected, the $r_{\textrm{zeroth}} \propto h$ and the $r_{\textrm{first}} \propto h^2$, which demonstrates the correct computation of the gradient by JAM-FEM.

\subsection{Topology optimization}

\begin{figure}[H] 
\centering
\includegraphics[scale=0.65]{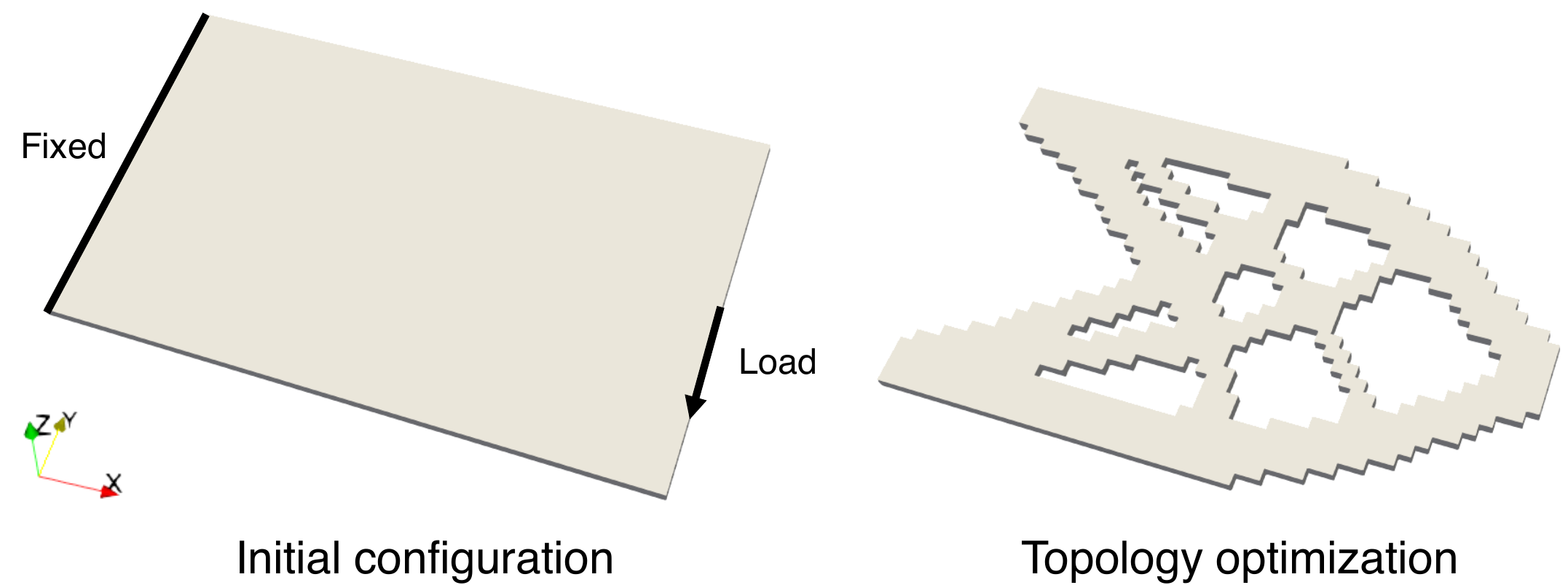}
\caption{Topology optimization of a thin plate.}
\label{Fig:plate}
\end{figure}

As the second numerical example of PDE-CO, we consider topology optimization~\cite{bendsoe2003topology}, an important field that is well-received and developed.
We first study compliance minimization of a thin plate made of a hyperelastic material, as shown in the left panel of Fig.~\ref{Fig:plate}.
Following the classic Solid Isotropic Material with Penalization (SIMP)~\cite{bendsoe2003topology} method, the governing PDE is 
\begin{align} \label{Eq:top_strong}
    -\nabla \cdot (\theta^p \Pb) = \boldsymbol{0} & \quad \textrm{in}  \, \, \Omega, \nonumber \\
    \ub = \boldsymbol{0} &  \quad\textrm{on} \, \, \Gamma_D,  \nonumber \\
    \theta^p \Pb \cdot \nb =  \tb & \quad \textrm{on} \, \, \Gamma_N,
\end{align}
which is similar to Eq.~(\ref{Eq:hyperelastic_strong}), except that $\theta(\x) \in [0, 1]$ is the continuous design density field and $p$ is the penalty exponent.
The weak form states that find the solution $\ub$ so that for any test function $\vb$ we have
\begin{align} \label{Eq:top_weak}
\int_{\Omega}  \theta^p \Pb :  \nabla \vb \textrm{ d} \Omega - \int_{\Gamma_N} \tb \cdot  \vb \textrm{ d} \Gamma = 0.
\end{align}
The compliance minimization problem states that 
\begin{align} \label{Eq:top_inverse}
    \nonumber \min_{\Ub\in\sR^{N}, \thetab\in\sR^{M}}  \int_{\Gamma_N} \ub^h \cdot \tb  \\
    \textrm{s.t.} \quad \Cb(\Ub, \thetab) = \textbf{0}, 
\end{align}
where $\ub^h(\x) = \sum_k \Ub[k] \phib_k(\x)$  is the finite element solution field constructed with $\Ub$, $\thetab$ is the discretized version of $\theta$, and the constraint equation $\Cb(\Ub, \thetab) = \textbf{0}$ matches the discretized version of Eq.~(\ref{Eq:top_weak}).
With \texttt{JAX-FEM}, we bypass the need to further perform sensitivity analysis, which is usually required in topology optimization.
The sensitivity information is computed by the program automatically.
The optimized topological structure of this thin plate is shown in the right panel of Fig.~\ref{Fig:plate}.
The optimizer used in this example is the method of moving asymptotes (MMA)~\cite{svanberg1987method}.
We have also passed a constraint that requires to only use 50$\%$ of the material, i.e., $\int_{\Omega} \theta \textrm{d}\Omega/\int_{\Omega} \textrm{d}\Omega =0.5$ to the optimizer.
To avoid checkerboard patterns~\cite{sigmund1998numerical}, a convolution filter is used to blur the calculated sensitivities.
The implementation of MMA in $\texttt{Python}$ is borrowed from the work of Chandrasekhar et al.~\cite{chandrasekhar2021auto}.

\begin{figure}[H] 
\centering
\includegraphics[scale=0.45]{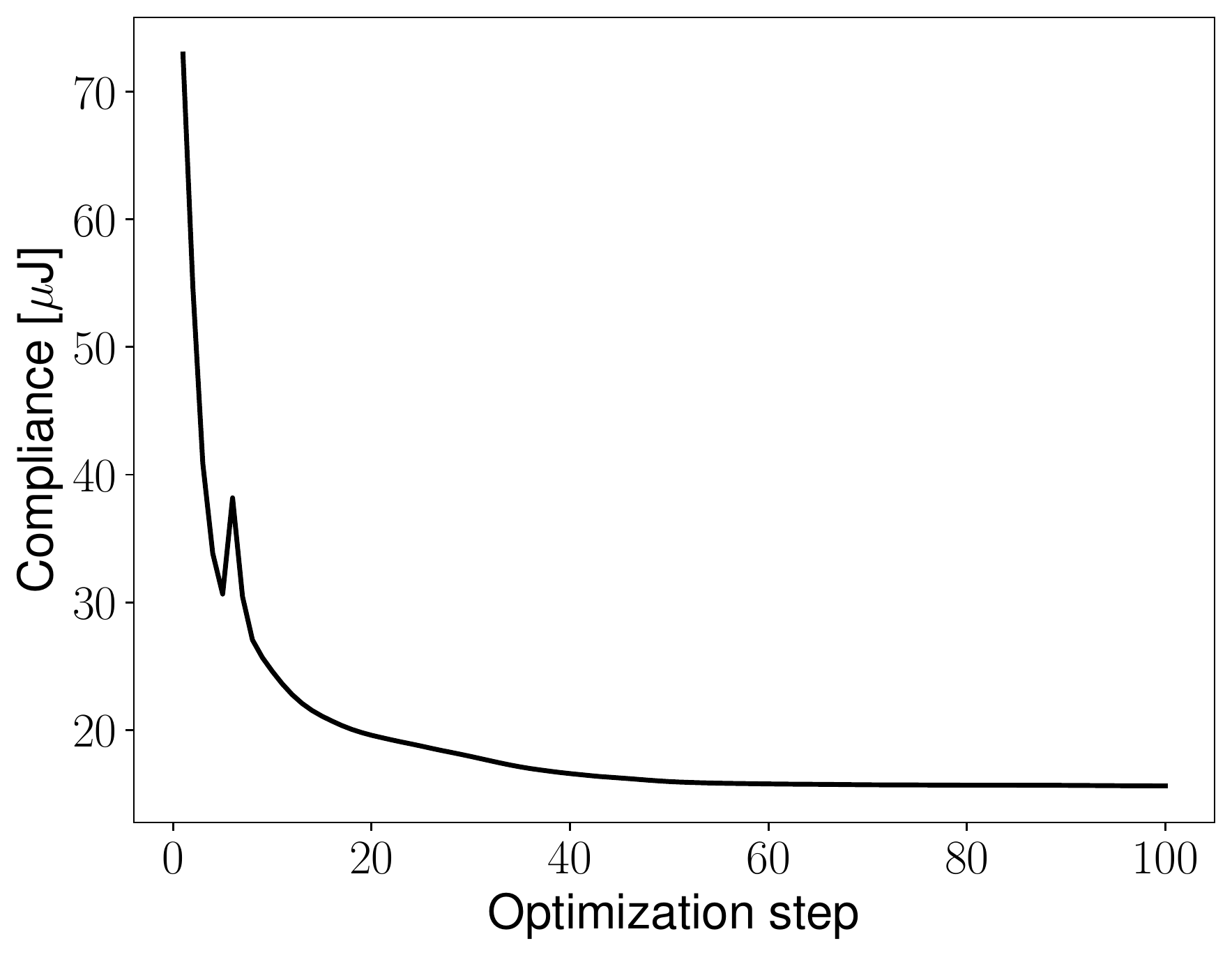}
\caption{Objective value versus optimization step for 
the thin plate example.}
\label{Fig:top_opt_plate}
\end{figure}

The compliance versus optimization step is shown in Fig.~\ref{Fig:top_opt_plate}, where the final structure has a compliance value of 15.63 $\mu$J.
As a reminder, the original solid plate with full material has the compliance to be 9.46 $\mu$J.

\begin{figure}[H] 
\centering
\includegraphics[scale=0.75]{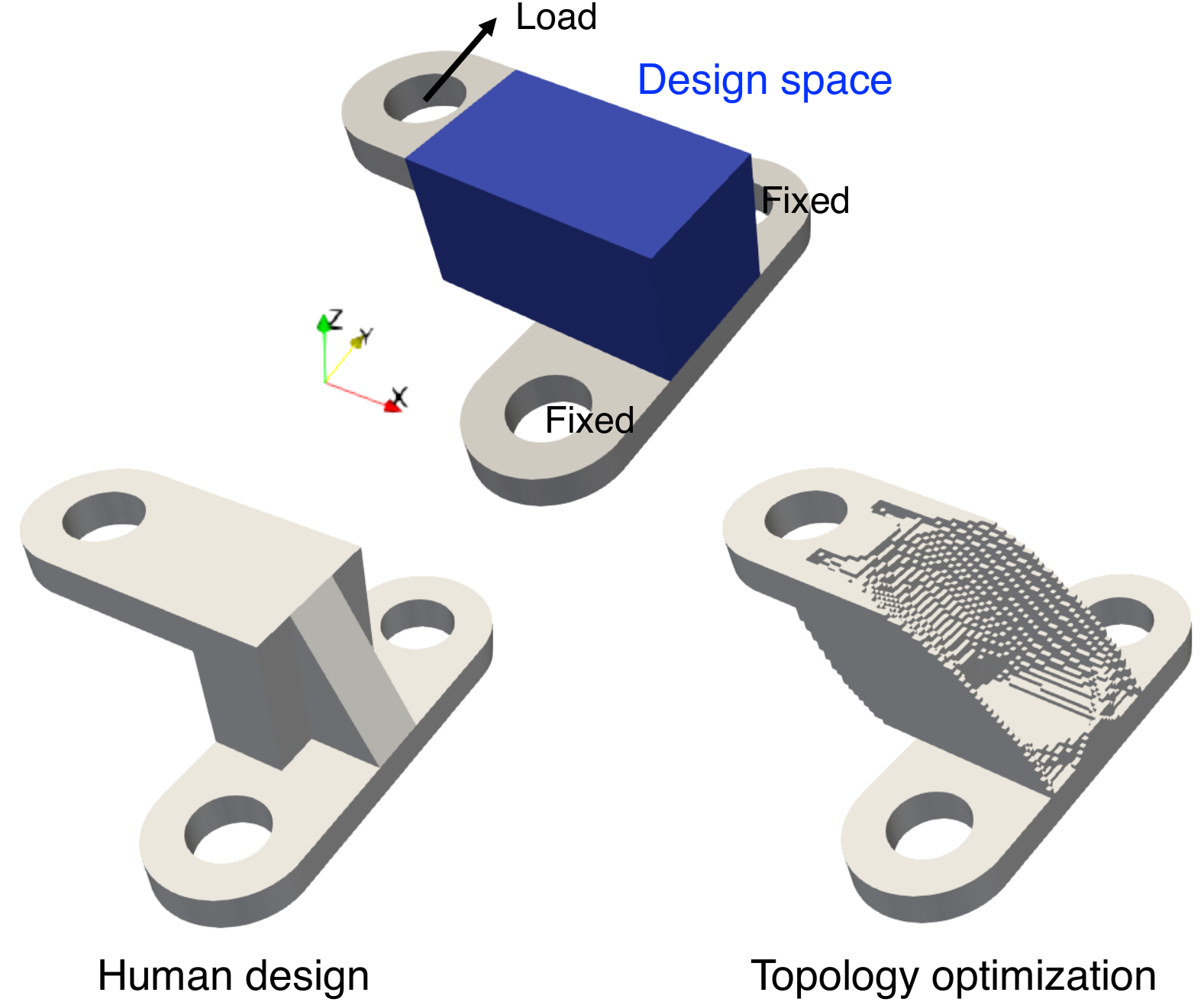}
\caption{Topology optimization of a three-hole bracket. As shown in the top panel, fixed boundary conditions are applied on the inner walls of the two holes, and uniform loading condition along the positive $y$-axis is applied on the inner wall of the upper hole. The lower left panel shows a human design while the lower right panel shows the result of topology optimization that uses the same amount of material.}
\label{Fig:bracket}
\end{figure}

\begin{figure}[H] 
\centering
\includegraphics[scale=0.8]{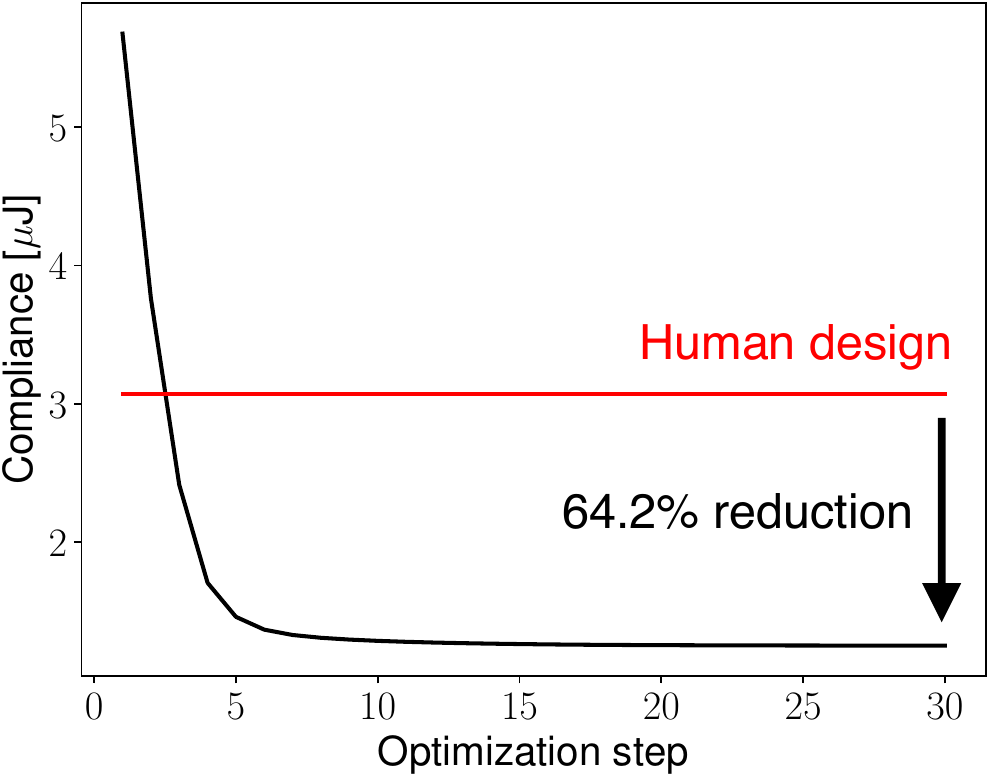}
\caption{Objective value versus optimization step for the bracket example.}
\label{Fig:top_opt_freecad}
\end{figure}

As a more realistic case with more complex geometry, we use topology optimization for the lightweight design of a bracket with three screw holes.
The bracket is assumed to be made of a linear elastic material and the boundary conditions are shown in the upper panel of Fig.~\ref{Fig:bracket}.
We limit the design space to the blue box region (see Fig.~\ref{Fig:bracket}), and prohibit any material change outside of the blue box.
A reasonable human design is shown in the lower left panel that uses 45$\%$ material of the blue box region, and the compliance is 3.07 $\mu$J.
The deign out of topology optimization that uses the same amount of material is shown in the lower right panel of Fig.~\ref{Fig:bracket}, whose compliance is 1.10 $\mu$J, achieving a reduction of 64.2$\%$ compared with the human design.
The optimization iterations are shown in Fig.~\ref{Fig:top_opt_freecad}  where the compliance value is plotted against the optimization step.

\section{Integration with machine learning}
\label{Sec:ml}

In this section, we show one numerical example of solving a data-driven multi-scale computation problem.
\texttt{JAX-FEM} provides an all-in-one solution to problems of this kind.
 
\subsection{Data-driven homogenization of composite material}

\begin{figure}[H] 
\centering
\includegraphics[scale=0.5]{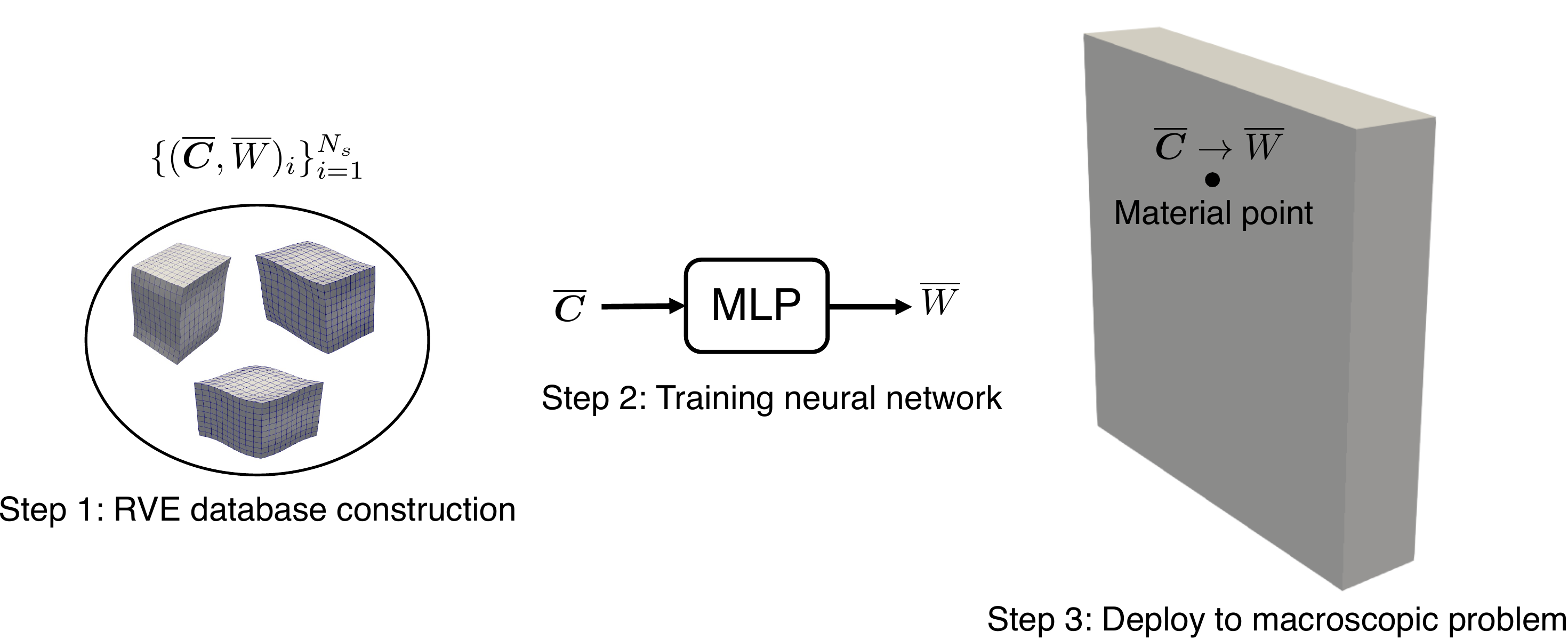}
\caption{Workflow of the data-driven multi-scale computational scheme. ``MLP'' stands for multi-layer perceptron. }
\label{Fig:overview}
\end{figure}

\begin{figure}[H] 
\centering
\includegraphics[scale=0.55]{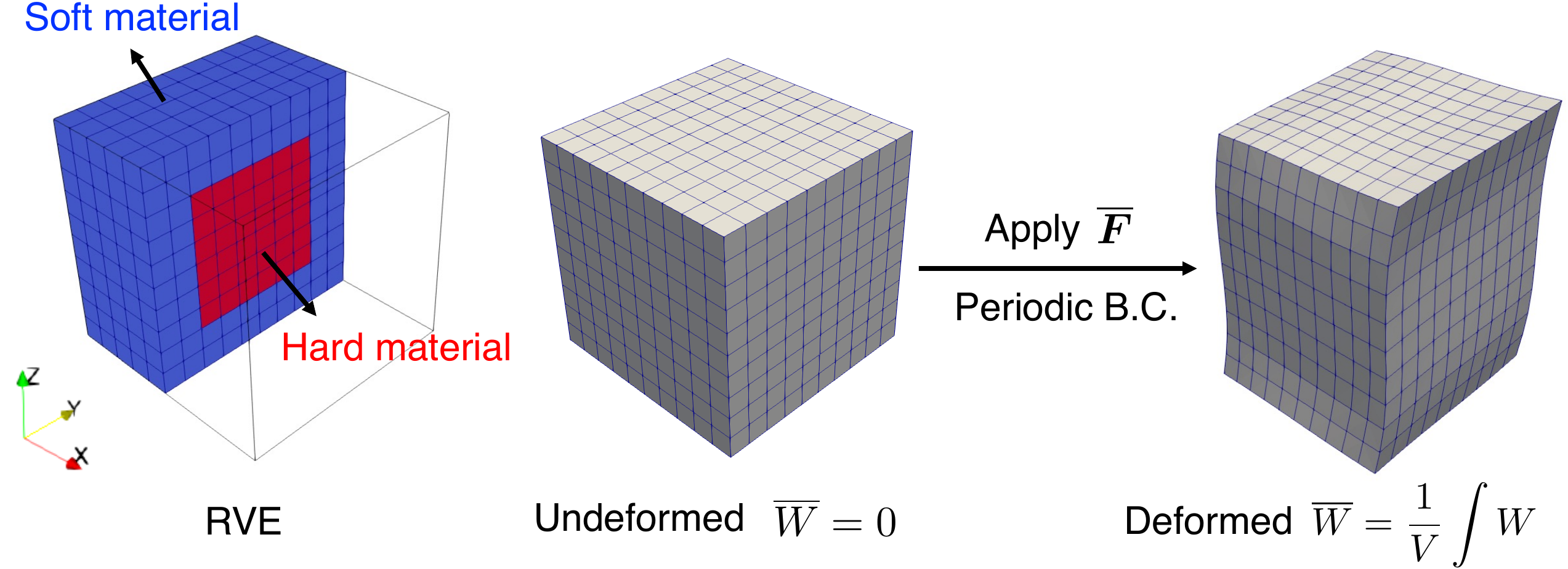}
\caption{The RVE computation mechanism. The left panel shows half of the RVE (the other half is symmetric). The middle and right panels show the deformation of an RVE subject to a macroscopic deformation gradient $\overline{\Fb}$ condition and periodic boundary conditions.}
\label{Fig:RVE}
\end{figure}

We consider a composite material whose representative volume element (RVE) is a 1\,mm cube of the mixture of a soft material ($E=$100 MPa; $\nu=0.4$) and a hard material ($E=$1000 MPa; $\nu=0.3$), as shown in the left panel of Fig.~\ref{Fig:RVE}.
Both soft and hard materials are assumed to be nearly incompressible neo-Hookean solids (similar to Eq.~(\ref{Eq:neo-hookean})).
The multi-scale computational scheme follows our previous work~\cite{xue2020data} on data-driven homogenization of mechanical meta-materials. 
The basic workflow involves three major steps: 
\begin{enumerate}
    \item Performing RVE-level FEM computations and collecting data;
    \item Training the neural network that represents the homogenized constitutive relationship; 
    \item Deploying the trained model to solve a macroscopic problem.
\end{enumerate}
The three steps are described in Fig.~\ref{Fig:overview}, where $\overline{\Cb}$ is the macroscopic right Cauchy-Green tensor and $\overline{W}$ is the macroscopic strain energy density function. 
For usual workflows, one needs to conduct step 1 with FEM software such as~\texttt{Abaqus}, conduct step 2 with a machine learning library such as~\texttt{PyTorch}~\cite{paszke2019pytorch} and then perhaps most tediously in step 3 get back to the FEM software so that the trained neural network parameters are hard-coded in the user-defined material model program.
In our work, all these three steps are performed in \texttt{JAX-FEM}, which is efficient and convenient.
In the next three subsections, we introduce the three steps in more detail.

\subsubsection{RVE database construction}

For each RVE problem, we prescribe a macroscopic deformation gradient $\overline{\Fb}$ and use FEM to solve for the displacement field $\ub(\X)$, where $\X$ is the material point.
As for the boundary condition, the solution $\ub(\X)$ is decomposed into a macroscopic (overall) part~$\overline{\ub} = (\overline{\Fb} - \Ib )\cdot \X$ and a microscopic (fluctuating) part~$\ub^{\star}$ such that
\begin{align}
\label{eqn:u-u*}
\ub(\X) = (\overline{\Fb} - \Ib )\cdot \X + \ub^{\star}(\X),
\end{align} 
where the fluctuating part $\ub^\star$ satisfies periodic boundary conditions.
Then the macroscopic energy density $\overline{W}$ is calculated as a volume averaged quantity:
\begin{align}
    \overline{W} = \frac{1}{V} \int W,
\end{align}
where $W$ is the strain energy density that depends on $\ub$ and $V$ is the RVE volume.
Due to material frame indifference~\cite{holzapfel2002nonlinear}, $\overline{W}$ is a function of the macroscopic right Cauchy-Green tensor $\overline{\Cb}=\overline{\Fb}^\top\overline{\Fb}$ and that $\overline{W}(\overline{\Cb})$ is the nonlinear constitutive relation we want to approximate with a data-driven approach.
The RVE computation is repeated with different $\overline{\Cb}$ so that a database $\{(\overline{\Cb}, \overline{W})\}_{i=1}^{N_s}$ is constructed for supervised learning.
Each data point contains the feature vector $\overline{\Cb}$ and the scalar label $\overline{W}$.
Sobol sequence~\cite{sobol1967distribution} method is used to generate around 1000 random data points for training.

\subsubsection{Training, validation and testing}

\begin{figure}[H] 
\centering
\includegraphics[scale=0.6]{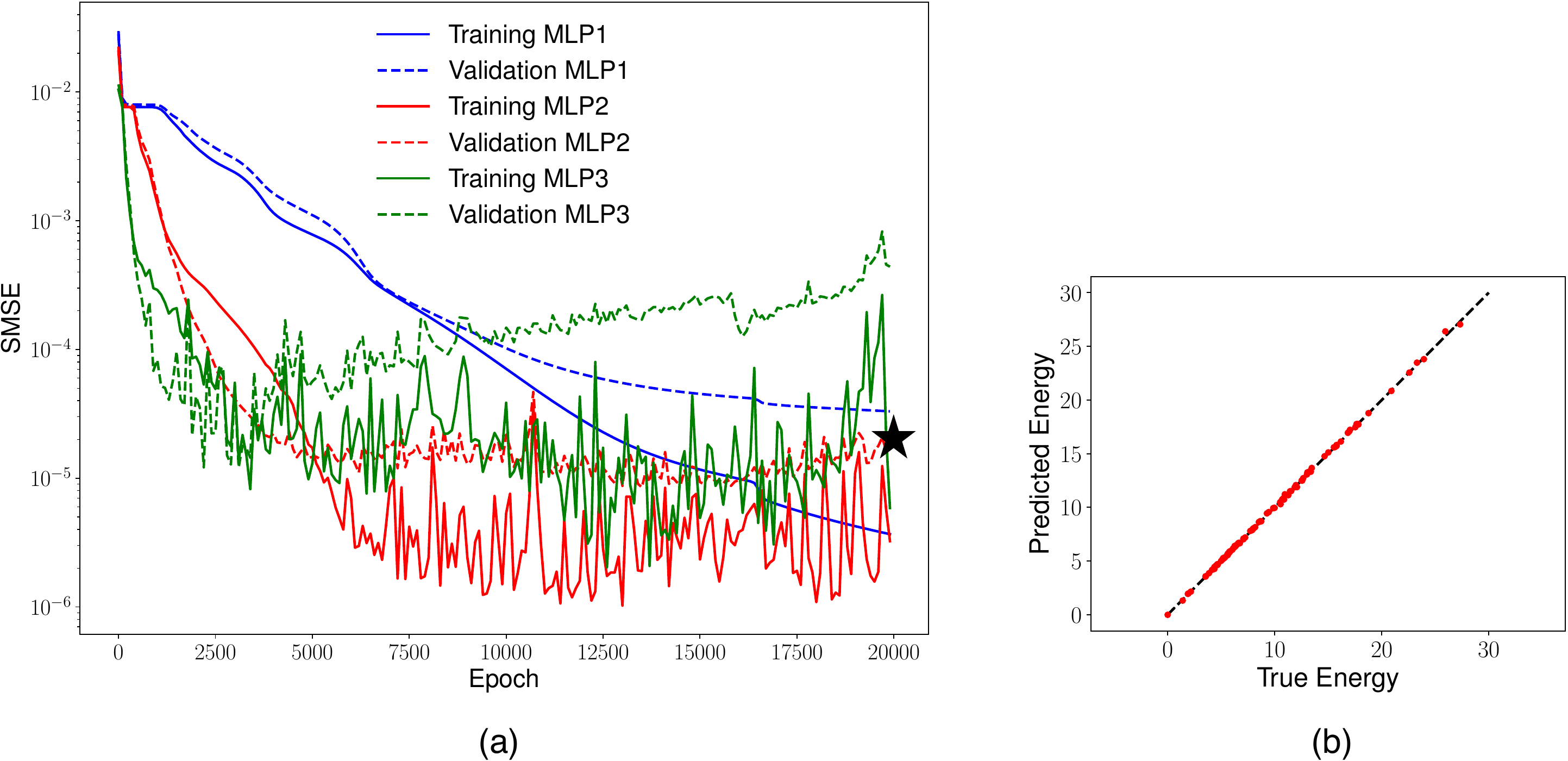}
\caption{Training, validation, and testing of the models. In (a), training and validation errors are shown for all three MLPs. The black star shows the lowest validation error. In (b), we show the test result for the selected MLP2.}
\label{Fig:training}
\end{figure}

We train multi-layer perceptron (MLP)~\cite{bishop2006pattern} for an approximate surrogate model such that $\overline{W}_{\textrm{MLP}}(\overline{\Cb})\approx\overline{W}(\overline{\Cb})$, as shown in the middle panel of Fig.~\ref{Fig:overview}.
The data set is split with an 8:1:1 ratio for training, model validation, and testing.
Three MLPs with increasing model capability are considered: ``MLP1'' has 4 hidden layers and each hidden has 32 neurons; ``MLP2'' has 8 hidden layers and 64 layer width; ``MLP3'' has 16 hidden layers and 128 layer width.
The activation function is set to be the hyperbolic tangent function.
The scaled mean square error (SMSE) is used as the criterion to select the optimal model, which is defined as
\begin{align}
    \textrm{SMSE} = \frac{1}{N_s} \sum_{i=1}^{N_{s}} (\bar{y}_{\textrm{true}} - \bar{y}_{\textrm{pred}})^2 \quad \textrm{with} \quad \bar{y} := \frac{y - y_{\textrm{min}}}{y_{\textrm{max}} - y_{\textrm{min}}},
\end{align}
where $N_s$ is the number of samples considered, $\bar{y}_{\textrm{true}}$ is the scaled true output, $\bar{y}_{\textrm{pred}}$ is the scaled predicted output, and  $(y_{\textrm{min}}, y_{\textrm{max}})$ are the lower and upper bounds in the training data.
We report training and validation SMSEs for the three MLPs and Fig.~\ref{Fig:training} (a).
As shown, MLP2 has the best validation SMSE and is selected for deployment.
The test result of MLP2 is shown in Fig.~\ref{Fig:training} (b) where the predicted macroscopic strain energy density values match the true values well.

\subsubsection{Macroscopic problem}

\begin{figure}[H] 
\centering
\includegraphics[scale=0.65]{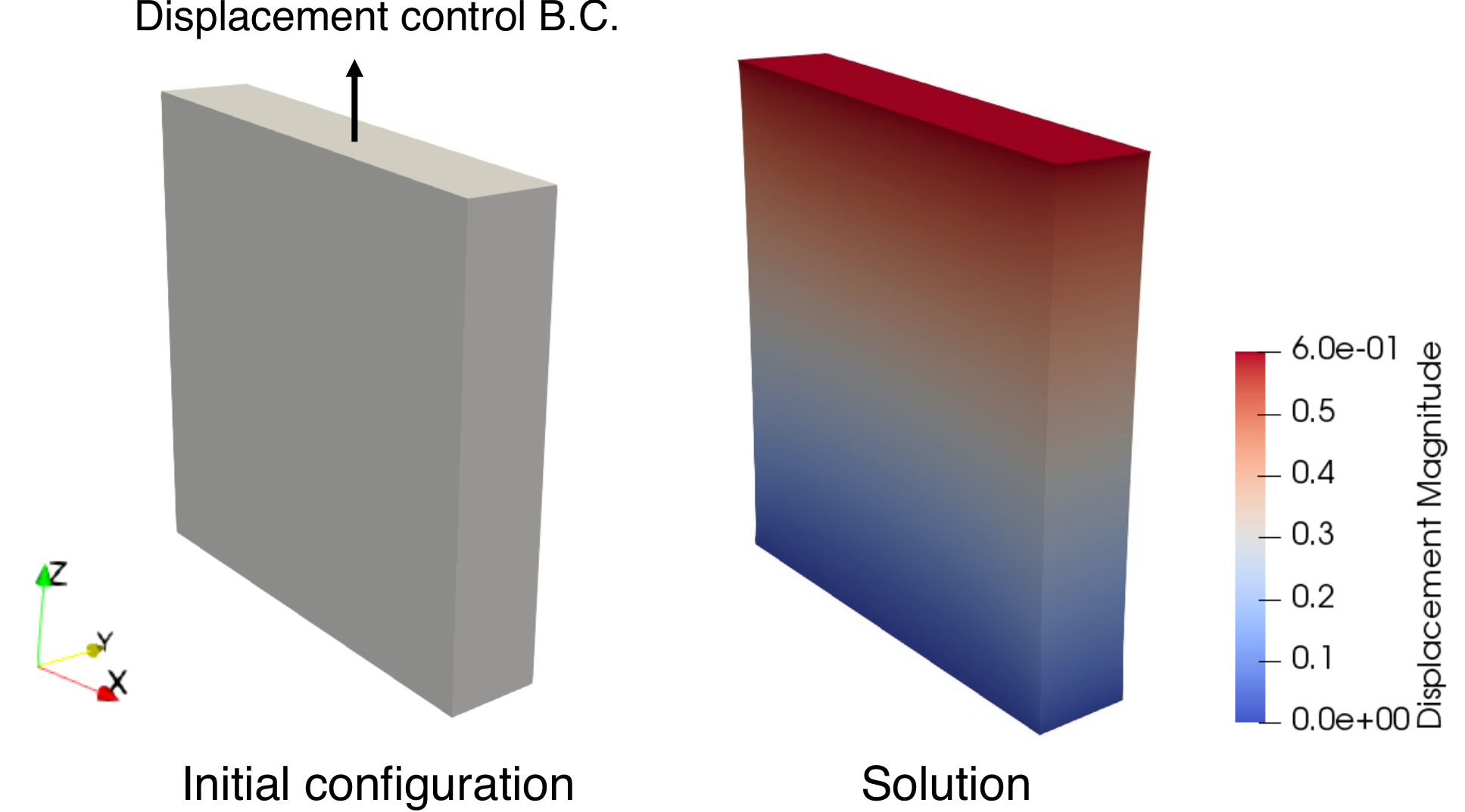}
\caption{Problem setup. The left panel shows the boundary condition such that the bottom surface is fixed and the top surface is subject to a prescribed $y$-axis displacement. The right penal shows the deformed configuration.}
\label{Fig:deploy_setup}
\end{figure}

The trained MLP2 represents the homogenized constitutive relationship and is deployed to solve a macroscopic problem.
We consider a uni-axial tensile loading on a $10\times 2 \times 10$ mm$^3$ (consisting of $10\times 2 \times 10$ RVEs) sample. 
The problem setup is shown in Fig.~\ref{Fig:deploy_setup}, where quasi-static loading up to 1 mm (10$\%$ of the $y$-axis sample size) is applied.

\begin{figure}[H] 
\centering
\includegraphics[scale=0.65]{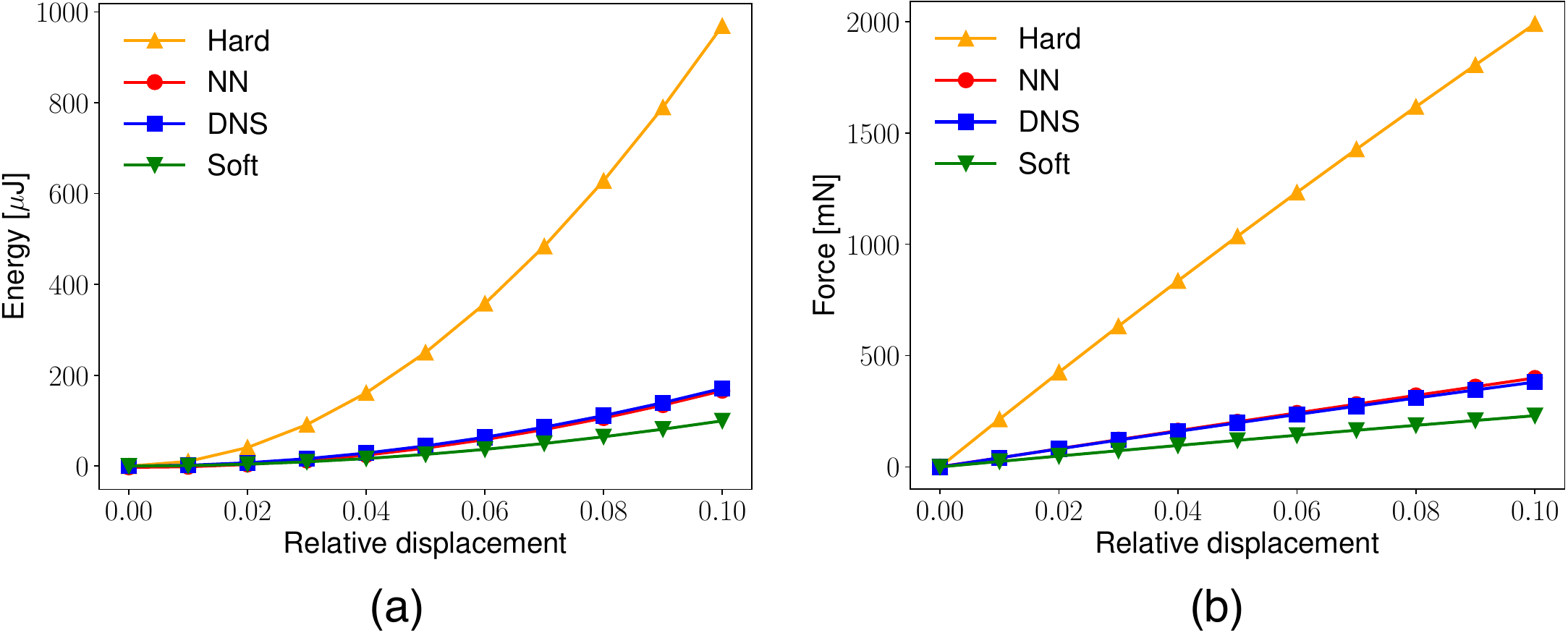}
\caption{Comparison between DNS and NN results. In (a), the total energy is shown with respect to relative displacement up to $10\%$. In (b), force on the top surface is plotted.}
\label{Fig:deploy_plot}
\end{figure}

We use both direct numerical simulation (DNS) and neural network (NN) surrogate models to solve the problem.
The total elastic energy stored in the sample as well as the tensile force applied on the top surface is plotted in Figs.~\ref{Fig:deploy_plot} (a) and (b), respectively.
Mechanical responses of bulk materials that are made solely by the hard and soft materials are also shown in the figure for reference. 
We observe good agreements between DNS and NN results, showing the effectiveness of the proposed data-driven multi-scale computational approach.
DNS uses 200,000 FEM cells while the NN-based model only uses 25,000 FEM cells, being more computationally efficient. 

\section{Conclusions and future work}
\label{Sec:conclusion}

We proposed and shared with the community an open-source FEM library \texttt{JAX-FEM}, a fundamental tool that efficiently solves the forward/inverse problems and facilitates research in data-driven computational mechanics.
The software can be more powerful and we list the following considerations for possible future improvement:
\begin{enumerate}
    \item The basic FEM toolkit needs to be more complete, e.g., richer element types, support of triangular/tetrahedron mesh, etc.
    \item The current linear solver is the biconjugate gradient stabilized method~\cite{van1992bi} with the simplest  Jacobi preconditioner. 
    A better linear solver with a suitable preconditioner can greatly improve the performance of \texttt{JAX-FEM}. 
    Interfacing with powerful external linear solvers like \texttt{PETSc}~\cite{balay2019petsc} is in progress.
    \item Inverse problems considered in this paper are all deterministic. 
    We plan to solve Bayesian inverse problems~\cite{stuart2010inverse} and consider uncertainty quantification in the future.
    \item The largest problem we can solve with a single 48 GB memory NVIDIA GPU is around 10 million DOF, otherwise the memory is insufficient.
    A multi-GPU version of \texttt{JAX-FEM} is highly-desired and is our future research goal. 
\end{enumerate}




\bibliographystyle{unsrt}
\bibliography{refs}

\begin{thebibliography}{10}

\bibitem{kamensky2019tigar}
David Kamensky and Yuri Bazilevs.
\newblock tigar: Automating isogeometric analysis with fenics.
\newblock {\em Computer Methods in Applied Mechanics and Engineering},
  344:477--498, 2019.

\bibitem{hughes2012finite}
Thomas~JR Hughes.
\newblock {\em The finite element method: linear static and dynamic finite
  element analysis}.
\newblock Courier Corporation, 2012.

\bibitem{jax2018github}
James Bradbury, Roy Frostig, Peter Hawkins, Matthew~James Johnson, Chris Leary,
  Dougal Maclaurin, George Necula, Adam Paszke, Jake Vander{P}las, Skye
  Wanderman-{M}ilne, and Qiao Zhang.
\newblock {JAX}: composable transformations of {P}ython+{N}um{P}y programs,
  2018.

\bibitem{kochkov2021machine}
Dmitrii Kochkov, Jamie~A Smith, Ayya Alieva, Qing Wang, Michael~P Brenner, and
  Stephan Hoyer.
\newblock Machine learning--accelerated computational fluid dynamics.
\newblock {\em Proceedings of the National Academy of Sciences},
  118(21):e2101784118, 2021.

\bibitem{bezgin2023jax}
Deniz~A Bezgin, Aaron~B Buhendwa, and Nikolaus~A Adams.
\newblock Jax-fluids: A fully-differentiable high-order computational fluid
  dynamics solver for compressible two-phase flows.
\newblock {\em Computer Physics Communications}, 282:108527, 2023.

\bibitem{xue2022learning}
Tianju Xue, Sigrid Adriaenssens, and Sheng Mao.
\newblock Learning the nonlinear dynamics of soft mechanical metamaterials with
  graph networks.
\newblock {\em arXiv preprint arXiv:2202.13775}, 2022.

\bibitem{schoenholz2020jax}
Samuel Schoenholz and Ekin~Dogus Cubuk.
\newblock Jax md: a framework for differentiable physics.
\newblock {\em Advances in Neural Information Processing Systems}, 33, 2020.

\bibitem{xue2022physics}
Tianju Xue, Zhengtao Gan, Shuheng Liao, and Jian Cao.
\newblock Physics-embedded graph network for accelerating phase-field
  simulation of microstructure evolution in additive manufacturing.
\newblock {\em npj Computational Materials}, 8(1):1--13, 2022.

\bibitem{bangerth2007deal}
Wolfgang Bangerth, Ralf Hartmann, and Guido Kanschat.
\newblock deal. ii—a general-purpose object-oriented finite element library.
\newblock {\em ACM Transactions on Mathematical Software (TOMS)}, 33(4):24--es,
  2007.

\bibitem{griewank2008evaluating}
Andreas Griewank and Andrea Walther.
\newblock {\em Evaluating derivatives: principles and techniques of algorithmic
  differentiation}.
\newblock SIAM, 2008.

\bibitem{lecun2015deep}
Yann LeCun, Yoshua Bengio, and Geoffrey Hinton.
\newblock Deep learning.
\newblock {\em nature}, 521(7553):436--444, 2015.

\bibitem{vigliotti2021automatic}
Andrea Vigliotti and Ferdinando Auricchio.
\newblock Automatic differentiation for solid mechanics.
\newblock {\em Archives of Computational Methods in Engineering},
  28(3):875--895, 2021.

\bibitem{lindsay2021automatic}
Alexander Lindsay, Roy Stogner, Derek Gaston, Daniel Schwen, Christopher
  Matthews, Wen Jiang, Larry~K Aagesen, Robert Carlsen, Fande Kong, Andrew
  Slaughter, et~al.
\newblock Automatic differentiation in metaphysicl and its applications in
  moose.
\newblock {\em Nuclear Technology}, 207(7):905--922, 2021.

\bibitem{mozaffar4160375differentiable}
Mojtaba Mozaffar, Shuheng Liao, Jihoon Jeong, Tianju Xue, and Jian Cao.
\newblock Differentiable simulation for material thermal response design in
  additive manufacturing processes.
\newblock {\em Available at SSRN 4160375}.

\bibitem{rees2010optimal}
Tyrone Rees, H~Sue Dollar, and Andrew~J Wathen.
\newblock Optimal solvers for pde-constrained optimization.
\newblock {\em SIAM Journal on Scientific Computing}, 32(1):271--298, 2010.

\bibitem{van2005review}
F~Van~Keulen, RT~Haftka, and Na-Hyung Kim.
\newblock Review of options for structural design sensitivity analysis. part 1:
  Linear systems.
\newblock {\em Computer methods in applied mechanics and engineering},
  194(30-33):3213--3243, 2005.

\bibitem{errico1997adjoint}
Ronald~M Errico.
\newblock What is an adjoint model?
\newblock {\em Bulletin of the American Meteorological Society},
  78(11):2577--2592, 1997.

\bibitem{cao2003adjoint}
Yang Cao, Shengtai Li, Linda Petzold, and Radu Serban.
\newblock Adjoint sensitivity analysis for differential-algebraic equations:
  The adjoint {DAE} system and its numerical solution.
\newblock {\em SIAM Journal on Scientific Computing}, 24(3):1076--1089, 2003.

\bibitem{kanno2021kernel}
Yoshihiro Kanno.
\newblock A kernel method for learning constitutive relation in data-driven
  computational elasticity.
\newblock {\em Japan Journal of Industrial and Applied Mathematics},
  38(1):39--77, 2021.

\bibitem{mozaffar2019deep}
M~Mozaffar, R~Bostanabad, W~Chen, K~Ehmann, Jian Cao, and MA~Bessa.
\newblock Deep learning predicts path-dependent plasticity.
\newblock {\em Proceedings of the National Academy of Sciences},
  116(52):26414--26420, 2019.

\bibitem{xu2021learning}
Kailai Xu, Alexandre~M Tartakovsky, Jeff Burghardt, and Eric Darve.
\newblock Learning viscoelasticity models from indirect data using deep neural
  networks.
\newblock {\em Computer Methods in Applied Mechanics and Engineering},
  387:114124, 2021.

\bibitem{logg2012automated}
Anders Logg, Kent-Andre Mardal, and Garth Wells.
\newblock {\em Automated solution of differential equations by the finite
  element method: The FEniCS book}, volume~84.
\newblock Springer Science \& Business Media, 2012.

\bibitem{van2011numpy}
Stefan Van Der~Walt, S~Chris Colbert, and Gael Varoquaux.
\newblock The numpy array: a structure for efficient numerical computation.
\newblock {\em Computing in science \& engineering}, 13(2):22--30, 2011.

\bibitem{harris2020array}
Charles~R Harris, K~Jarrod Millman, St{\'e}fan~J Van Der~Walt, Ralf Gommers,
  Pauli Virtanen, David Cournapeau, Eric Wieser, Julian Taylor, Sebastian Berg,
  Nathaniel~J Smith, et~al.
\newblock Array programming with numpy.
\newblock {\em Nature}, 585(7825):357--362, 2020.

\bibitem{ogden1997non}
Raymond~W Ogden.
\newblock {\em Non-linear elastic deformations}.
\newblock Courier Corporation, 1997.

\bibitem{simo2006computational}
Juan~C Simo and Thomas~JR Hughes.
\newblock {\em Computational inelasticity}, volume~7.
\newblock Springer Science \& Business Media, 2006.

\bibitem{Gooch2011}
Jan~W. Gooch.
\newblock {\em ASTM D638}, pages 51--51.
\newblock Springer New York, New York, NY, 2011.

\bibitem{betts2005discretize}
John~T Betts and Stephen~L Campbell.
\newblock Discretize then optimize.
\newblock {\em Mathematics for industry: challenges and frontiers}, pages
  140--157, 2005.

\bibitem{liu2019non}
Jun Liu and Zhu Wang.
\newblock Non-commutative discretize-then-optimize algorithms for elliptic
  pde-constrained optimal control problems.
\newblock {\em Journal of Computational and Applied Mathematics}, 362:596--613,
  2019.

\bibitem{MR0055409}
Walter Rudin.
\newblock {\em Principles of mathematical analysis}.
\newblock McGraw-Hill Book Company, Inc., New York-Toronto-London, 1953.

\bibitem{xu2022physics}
Kailai Xu and Eric Darve.
\newblock Physics constrained learning for data-driven inverse modeling from
  sparse observations.
\newblock {\em Journal of Computational Physics}, 453:110938, 2022.

\bibitem{blondel2021efficient}
Mathieu Blondel, Quentin Berthet, Marco Cuturi, Roy Frostig, Stephan Hoyer,
  Felipe Llinares-L{\'o}pez, Fabian Pedregosa, and Jean-Philippe Vert.
\newblock Efficient and modular implicit differentiation.
\newblock {\em arXiv preprint arXiv:2105.15183}, 2021.

\bibitem{byrd1995limited}
Richard~H Byrd, Peihuang Lu, Jorge Nocedal, and Ciyou Zhu.
\newblock A limited memory algorithm for bound constrained optimization.
\newblock {\em SIAM Journal on scientific computing}, 16(5):1190--1208, 1995.

\bibitem{virtanen2020scipy}
Pauli Virtanen, Ralf Gommers, Travis~E Oliphant, Matt Haberland, Tyler Reddy,
  David Cournapeau, Evgeni Burovski, Pearu Peterson, Warren Weckesser, Jonathan
  Bright, et~al.
\newblock Scipy 1.0: fundamental algorithms for scientific computing in python.
\newblock {\em Nature methods}, 17(3):261--272, 2020.

\bibitem{niewiarowski2020adjoint}
Alexander Niewiarowski, Sigrid Adriaenssens, and Ruy~Marcelo Pauletti.
\newblock Adjoint optimization of pressurized membrane structures using
  automatic differentiation tools.
\newblock {\em Computer Methods in Applied Mechanics and Engineering},
  372:113393, 2020.

\bibitem{xue2022mapped}
Tianju Xue and Sheng Mao.
\newblock Mapped shape optimization method for the rational design of cellular
  mechanical metamaterials under large deformation.
\newblock {\em International Journal for Numerical Methods in Engineering},
  123(10):2357--2380, 2022.

\bibitem{bendsoe2003topology}
Martin~Philip Bendsoe and Ole Sigmund.
\newblock {\em Topology optimization: theory, methods, and applications}.
\newblock Springer Science \& Business Media, 2003.

\bibitem{svanberg1987method}
Krister Svanberg.
\newblock The method of moving asymptotes—a new method for structural
  optimization.
\newblock {\em International journal for numerical methods in engineering},
  24(2):359--373, 1987.

\bibitem{sigmund1998numerical}
Ole Sigmund and Joakim Petersson.
\newblock Numerical instabilities in topology optimization: a survey on
  procedures dealing with checkerboards, mesh-dependencies and local minima.
\newblock {\em Structural optimization}, 16(1):68--75, 1998.

\bibitem{chandrasekhar2021auto}
Aaditya Chandrasekhar, Saketh Sridhara, and Krishnan Suresh.
\newblock Auto: a framework for automatic differentiation in topology
  optimization.
\newblock {\em Structural and Multidisciplinary Optimization},
  64(6):4355--4365, 2021.

\bibitem{xue2020data}
Tianju Xue, Alex Beatson, Maurizio Chiaramonte, Geoffrey Roeder, Jordan~T Ash,
  Yigit Menguc, Sigrid Adriaenssens, Ryan~P Adams, and Sheng Mao.
\newblock A data-driven computational scheme for the nonlinear mechanical
  properties of cellular mechanical metamaterials under large deformation.
\newblock {\em Soft matter}, 16(32):7524--7534, 2020.

\bibitem{paszke2019pytorch}
Adam Paszke, Sam Gross, Francisco Massa, Adam Lerer, James Bradbury, Gregory
  Chanan, Trevor Killeen, Zeming Lin, Natalia Gimelshein, Luca Antiga, et~al.
\newblock Pytorch: An imperative style, high-performance deep learning library.
\newblock {\em Advances in neural information processing systems}, 32, 2019.

\bibitem{holzapfel2002nonlinear}
Gerhard~A Holzapfel.
\newblock Nonlinear solid mechanics: a continuum approach for engineering
  science.
\newblock {\em Meccanica}, 37(4):489--490, 2002.

\bibitem{sobol1967distribution}
Il'ya~Meerovich Sobol'.
\newblock On the distribution of points in a cube and the approximate
  evaluation of integrals.
\newblock {\em Zhurnal Vychislitel'noi Matematiki i Matematicheskoi Fiziki},
  7(4):784--802, 1967.

\bibitem{bishop2006pattern}
Christopher~M Bishop and Nasser~M Nasrabadi.
\newblock {\em Pattern recognition and machine learning}, volume~4.
\newblock Springer, 2006.

\bibitem{van1992bi}
Henk~A Van~der Vorst.
\newblock Bi-cgstab: A fast and smoothly converging variant of bi-cg for the
  solution of nonsymmetric linear systems.
\newblock {\em SIAM Journal on scientific and Statistical Computing},
  13(2):631--644, 1992.

\bibitem{balay2019petsc}
Satish Balay, Shrirang Abhyankar, Mark Adams, Jed Brown, Peter Brune, Kris
  Buschelman, Lisandro Dalcin, Alp Dener, Victor Eijkhout, W~Gropp, et~al.
\newblock Petsc users manual.
\newblock 2019.

\bibitem{stuart2010inverse}
Andrew~M Stuart.
\newblock Inverse problems: a bayesian perspective.
\newblock {\em Acta numerica}, 19:451--559, 2010.

\end{thebibliography}


\end{document}